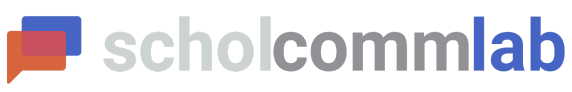

# Estimating global article processing charges paid to 14 publishers for open access between 2019 and 2025

Lisa Matthias[1, 2,*], Eric Schares[2,3], Juan Pablo Alperin[2,4],
Leigh-Ann Butler[2,5], Sherry Kuang[2,6], Nina Schönfelder[7] & Stefanie Haustein[2,6,8]

[1] Berlin School of Library and Information Science, Humboldt-Universität zu Berlin, Berlin (Germany)
[2] Scholarly Communications Lab, Ottawa/Vancouver (Canada)
[3] University Library, Iowa State University, Ames (USA)
[4] Simon Fraser University, Vancouver (Canada)
[5] University of Ottawa Library, Ottawa (Canada)
[6] School of Information Studies, University of Ottawa, Ottawa (Canada)
[7] University Library, Bielefeld University, Bielefeld (Germany)
[8] Centre interuniversitaire de recherche sur la science et la technologie (CIRST),
Université du Québec à Montréal, Montréal (Canada)

[*] *Corresponding author: l.a.matthia@gmail.com*

l.a.matthia@gmail.com - https://orcid.org/0000-0002-2612-2132
eschares@iastate.edu - https://orcid.org/0000-0002-6292-8221
juan@alperin.ca - https://orcid.org/0000-0002-9344-7439
leigh-Ann.Butler@uottawa.ca - https://orcid.org/0000-0002-3385-4729
skuan070@uottawa.ca - https://orcid.org/0009-0009-6982-3017
nina.schoenfelder@uni-bielefeld.de - https://orcid.org/0000-0001-5294-5354
stefanie.haustein@uottawa.ca - https://orcid.org/0000-0003-0157-1430

## Keywords
Article processing charges, scholarly publishing, open access, hybrid open access, gold open access, scholarly journals

# Abstract

This study presents estimates of the global expenditure on article processing charges (APCs) paid to 14 publishers for open access (OA) between 2019 and 2025. APCs are charged for publishing in fully OA journals (gold) and making individual articles OA in subscription journals (hybrid), but how much is paid, and for which articles, is not publicly known. We therefore curated an open dataset of publicly listed APC prices from 14 academic publishers (ACS, CUP, De Gruyter, EDP, Elsevier, Frontiers, IEEE, IOP, MDPI, OUP, PLOS, Sage, Springer Nature, and Wiley) and combined it with counts of OA articles from OpenAlex.

We estimate that $15.08 billion (in 2025 USD) was spent globally on APCs between 2019 and 2025. Adjusted for inflation, annual spending quadrupled from $0.9 billion in 2019 to $3.7 billion in 2025, with >85% concentrated among a few large publishers. Hybrid OA fees exceed gold fees, and the median fee paid is higher than the median price listed for both. Our approach addresses major limitations in previous efforts to estimate APC spending, offering much-needed

insight into an opaque aspect of scholarly publishing, especially as transformative agreements make it more challenging to understand the costs of publishing OA.

# 1. Introduction

In the twenty years since the advent of the open access (OA) movement, a myriad of publishing models have emerged as alternatives to well-established subscription models, contributing to the growing complexity of the OA publishing landscape. One dominant economic approach to OA for commercial publishers, widely known as the author-pays model, relies on the use of article processing charges (APCs), where publishers charge a fee for publishing their articles OA. APCs have proven lucrative for publishers (Butler et al., 2023; Jung et al., 2025; Shu & Larivière, 2024), but highly controversial for introducing barriers for researchers who struggle to pay several thousand dollars per article and for causing the diversion of resources away from research (Ayeni & Larivière, 2025; Halevi & Walsh, 2021; Nicholas et al., 2024). Although APCs are a popular revenue stream for OA publishing, especially among commercial publishers, there is much diversity in the OA publishing landscape, shaped by national approaches and factors such as funding availability (Kulczycki et al., 2026; van Bellen et al., 2025).

The "Big 5" commercial publishers—Elsevier, Sage, Springer Nature, Taylor & Francis, and Wiley—use APCs to supplement existing revenue strategies (Borrego, 2023; Butler, 2023; Butler et al., 2023), while other large publishers, such as Frontiers and MDPI, rely solely on APCs to generate revenue (Rodrigues et al., 2020). As the OA publishing landscape continues to experience large-scale growth (Piwowar et al., 2018; Shu & Larivière, 2024), the APC model is evolving in tandem with the support of funder and institutional OA policies and, more recently, through so-called "Transformative Agreements" (TA), also referred to as "Read and Publish" (R&P) agreements (Borrego et al., 2020). Their rapid growth since 2019 has been met with increased concerns surrounding institutional cost savings and an achievable transition from hybrid towards full OA transformation (Bakker et al., 2024; Baldwin & Cavanagh, 2024). The impact of TAs is contingent on a publisher's existing market presence, and favours large, commercial publishers through complex negotiations with longer terms and higher rates of renewal (Rothfritz et al., 2024; Schmal, 2024a). The growth of hybrid OA, especially owing to the increase of TAs, between 2019 and 2023 was similarly concentrated among large publishers such as Elsevier, Springer Nature, and Wiley (accounting for 62% of the hybrid articles, and 47% of the journals during the period), with large differences in adoption rates by country, with Europe and South Africa leading (Jahn, 2025a). By entrenching the largest publishers, these dynamics may narrow the space for emerging OA publishers, and reduce diversity in the publishing ecosystem (Schmal, 2024b). Measuring what institutions actually spend through these agreements is difficult since contract terms and invoicing data are rarely public, APC data is hard to monitor and connect to actual costs, and publisher reports are not normalized (Kemp & Skinner, 2024; Riegelman & Langham-Putrow, 2025). In response, recent work has turned to open data sources to estimate TA coverage (de Jonge et al., 2025; Jahn, 2025b), and to use such estimates to calculate per-article and total APC costs within and outside TAs (Neylon & Kramer, 2026).

The lack of systematic data on how much is spent on APCs limits the ability of institutions, funders, or consortia to make evidence-based decisions during negotiations with publishers or in the context of science policy. Studies analyzing APCs at the funder, country, institution, or disciplinary level have encountered several limitations. For example, the commonly used APC data from the Directory of Open Access Journals (DOAJ) does not include hybrid journals and lacks historical fees (Asai, 2023). There is no publicly available information on the prevalence of institutional discounts and waivers (Borrego et al., 2020; Riegelman & Langham-Putrow, 2025), and it is impossible to identify who ultimately pays the APC for a given publication, and through what mechanism (Asai, 2023; Borrego, 2023; Gallardo et al., 2024; Schönfelder, 2020). While institutions are better positioned to track their own expenditure on APCs, such information is not systematically tracked across institutions, leaving each to employ a range of methods to estimate their affiliated authors' spend, if they track it at all. In an effort to surface these data, the OpenAPC initiative in Germany (Pieper & Broschinski, 2018) has been crowdsourcing a collection of APCs actually paid by universities, funders, and research institutions from Europe and North America. These data are the most precise account of how much was actually paid for individual articles, but remains quite limited in scope: OpenAPC has records for only 287,105 paid APCs by 482 institutions from 2005 to 2026 (OpenAPC, 2026).

In the absence of comprehensive payment data, studies have estimated APC spend through several methods, such as: using an average APC (Schimmer et al., 2015; M. Smith et al., 2016; Swan & Houghton, 2012), applying the most current APC fee for a journal to older publications (Zhang et al., 2022), or relying on self-reported data via the OpenAPC initiative (Asai, 2024; Schönfelder, 2020). These approaches may serve the objectives of their studies, but risk significant over- or under-estimating. For example, Schönfelder and Tummes (2024) report that list-price APCs as recorded in the DOAJ overstate the actual paid APCs, as collected by the OpenAPC initiative, by 20% on average.

The approach taken here relies on publicly available list prices, which addresses many of the limitations of previous work. This study extends our earlier analysis (Haustein et al., 2024), which examined six publishers—Elsevier, Frontiers, MDPI, PLOS, Springer Nature, and Wiley—for 2019 to 2023. That analysis drew on the first version of our dataset (Butler et al., 2024b, 2024a). Here, we analyze an expanded second version of the APC dataset (Matthias et al., 2026b), created using a similar approach but covering 14 large publishers—American Chemical Society (ACS), Cambridge University Press (CUP), De Gruyter, EDP, Elsevier, Frontiers, IEEE, IOP, MDPI, Oxford University Press (OUP), PLOS, Sage, Springer Nature, and Wiley—and spans 2019 to 2025. We use this expanded dataset to estimate the global amount of APCs paid for OA publishing to these 14 publishers in recent years. More specifically, we seek to address the following research questions:

1. How much was paid in APCs to these six publishers for the 2019–2025 period?
    a. How did the estimated spend differ between gold and hybrid OA?
    b. How did the estimated spend differ between publishers?
    c. How did the estimated spend develop over time?

2. How do gold and hybrid APCs paid (article level) compare to the APCs listed (journal level)?

# 2. Methods

This study largely follows the methodology of estimating APC spend outlined in detail in Haustein et al. (2024). We compiled annual APC data into a new open dataset (Matthias et al., 2026b) and used the number of publications per journal per year from OpenAlex.

## 2.1 Annual APCs

The updated dataset  combines and standardizes data from the APC price lists of 14 large publishers. The dataset includes APC prices for 12,540 unique journals and 69,856 data points (i.e., journal-year combinations) spanning seven years (2019-2025). The dataset builds on 37 price lists used in the first version (Butler et al., 2024b), incorporating a further 59 APC price lists and websites and 21 journal lists to produce one coherent and reusable dataset. A detailed description of the dataset is given in Matthias et al. (2026a).

Coverage is not uniform across publishers and years (Figure 1). Eleven of the 14 publishers are covered for all seven years; De Gruyter and OUP first appear in 2021, and IOP is missing 2020. Three publishers—Frontiers, MDPI, and PLOS—publish gold OA only, so their records contain no hybrid fees. Sage is covered across all years, but its 2019 records capture gold OA only and its 2025 records hybrid OA only. Where a publisher-year is incompletely covered, this is flagged in the results below.

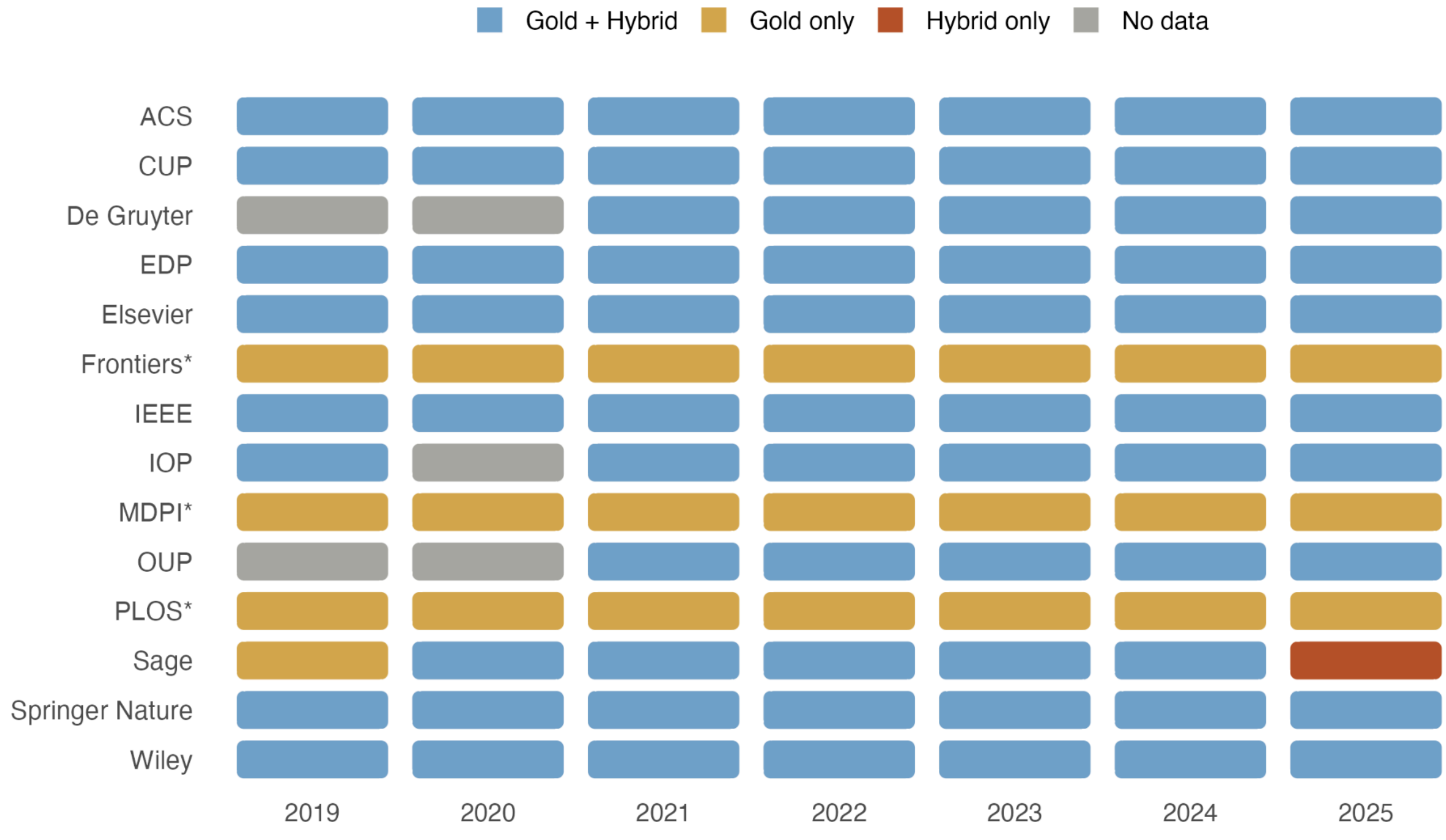

**Figure 1.** APC data coverage by publisher and year

Publisher price lists typically report at least one ISSN per journal, but do not always indicate whether it refers to the print or electronic version. We therefore do not make this distinction in the dataset, storing identifiers in two general ISSN columns (ISSN_1 and ISSN_2). A separate column (ISSN_L) holds the linking ISSN. Publishers provided APCs in multiple currencies. USD was the most frequently provided currency, at 94% of the 68,506 flat-rate journal-year combinations. The remaining 6% without USD original fees were converted from APCs provided in AUD, CAD, CHF, EUR, GBP, or JPY using the average annual exchange rate (Yahoo Finance, 2026).

Publishers structured their fees in several different ways, most commonly as flat rates (n = 68,506; 98%) but also through a range of other fee types. Springer Nature's rapid service fees (n = 38), which promise accelerated review and production for a fixed charge, and Elsevier's editorial processing charges (n = 1) each replaced a standard APC, so were treated as flat rates. We handled the remaining fee types according to how confidently we could determine the total charge. Where a flat rate carried a fixed and required secondary charge, such as publication charges (n = 4) and submission fees (n = 1), we added both and included the total. Where it carried a required page charge, which scales with article length (n = 135), we assumed a 10-page article to derive the total. Some journals levied a flat rate with overlength page charges (n = 395).

The number of pages at which the additional charge was introduced varied by journal, ranging from 4 to 10. In these cases we retained the flat rate but did not include any overlength surcharge. Where a flat rate carried a voluntary page charge (n = 35), we included only the flat rate, since we could not tell whether an author had chosen to pay it. A further set combined a flat rate with both a voluntary and an overlength charge (n = 236); as above, we retained the flat rate and excluded both additional charges.

Some journals charged page fees in place of a flat APC. For these (n = 22), we assumed a 10-page article and calculated the fee accordingly. Two further journals combined a page charge with a reviewing fee (n = 2), which we added together under the same 10-page assumption. A small set of IOP journals (n = 25) charged by "quanta"—units based on an article's word, figure, and table counts rather than page length. Drawing on an example acceptance report and personal correspondence with an *Astrophysical Journal* editor, we applied the middle fee tier for each year where available. When the APC was based on different charges for each type of quanta individually (n = 4), we used the acceptance report's breakdown to calculate the fee.

We excluded two fee types from our estimates because we could not determine what a typical author would pay: scaled APCs (n = 2) that vary with the corresponding author's economic context, offering discounts of 50% or 75% depending on where the author is based, and EDP's "Liberty APC" (n = 22), which lets authors pay whatever amount they choose. For an additional 432 journal-year combinations, no fee information could be determined, and these were excluded from the estimate.

## 2.2 Number of OA articles per journal per year

To estimate APC spend, we first determined the number of gold or hybrid OA publications per journal per year. This was done by querying the relevant journals, articles, and their access status in OpenAlex (Priem et al., 2022). Since OpenAlex is a new and evolving citation index with some known metadata limitations (Alperin et al., 2024; Culbert et al., 2025; Zhang et al., 2024), we compared publication counts per journal per year against Dimensions, when we first adopted this approach in 2024 (Haustein et al., 2024). We found the volume of articles by publisher and year to be within 2% of each other, giving us confidence that OpenAlex provides realistic counts for this purpose. We therefore report estimates based only on the openly available OpenAlex "Walden" data, March 2026 snapshot, accessed on Google Big Query through the ORION initiative (ORION, 2026).

### 2.2.1 Relevant document types

A main challenge is determining which articles can be plausibly considered to be "APC-able", that is, which of the documents published would have been subject to an APC. Generally, research and review article types fall into this category, while letters to the editor, corrections, viewpoints, and editorials do not. We identified documents in OpenAlex by filtering for those classified as *article* with a publication date between 2019 and 2025. Conference abstracts were detected by removing issues beginning with any letter or an underscore, or a first page beginning with an "S" or an underscore. A title search for any three numbers followed by the exact string

"pp. " removed book reviews. The remaining records were counted and the source matched to the ISSN-L in our price list dataset. The ISSN-L ("linking") was used to increase the chances of a match against all possibilities in the `ISSN` field, which can contain several variants for electronic, print, or historic ISSNs. The full SQL query is included in Supplement 1.

### 2.2.2 Open access status

Every journal-year combination in our dataset was classified as either gold or hybrid in its respective publisher price list. Gold journals are those for which all documents are APC-able OA, while hybrid journals are subscription-based and offer OA for individual articles upon payment of an APC. For gold journals, we did not use the OA status returned by OpenAlex due to known fluctuations and inaccuracies of the underlying Unpaywall algorithm (Jahn et al., 2021; Schares, 2023). Instead, we considered all documents returned by the SQL query in these journals as APC-able. The dataset contained 19,037 gold journal-year combinations, of which 18,647 had a known APC. Of these, 18,504 combinations could be matched to at least one article when matching by ISSN in OpenAlex, accounting for 4,214,160 gold OA articles published between 2019 and 2025 that we assume have been subjected to an APC.

For journals listed as hybrid, there were 50,796 journal-year combinations in the APC dataset, of which 50,747 contained APC information. When matching with OpenAlex using the journals' ISSN, we were able to identify at least one article for 49,825 journal-year combinations. In these journals, we only considered the portion of articles available as hybrid OA as having been subjected to an APC. We therefore relied on the OpenAlex OA status of each article to determine how many articles were APC-able.

When analyzing OA status of articles in hybrid journals, we noticed that some journals returned very high rates of hybrid OA (sometimes as high as 100%), contradicting reports of much lower hybrid rates around 4-10% (Jahn, 2025b; Jahn et al., 2022; Jahn & Haupka, 2022; Piwowar et al., 2019). Our manual investigation of high OA% Hybrid journals (>85%) revealed that these appear to be valid, and not an issue with delayed OA (i.e., subscription journals that make all their content available OA after a journal-specific embargo period) as reported previously.

After limiting hybrid articles to those with a hybrid open access designation, the resulting dataset contained 46,675 journal-year combinations with at least one APC-able article for 1,378,462 articles published in hybrid OA journals between 2019 and 2025 that we assume have been subjected to an APC.

## 2.3 Accounting for inflation

When reporting monetary trends over time it is useful to account for inflation. Previous studies estimating revenue and cost developments have converted prices from different datasets to the same base year value (Akbulut, 2026) or adjusted nominal APCs to real values using GDP deflators (Cumming, 2026). Inflation was particularly high in the seven-year period from 2019 to 2025: cumulative inflation reached 38.2% globally, 26.1% in the US, and 22.5% across Advanced Economies (International Monetary Fund, 2026).

Inflation adjustment was carried out on the list prices, adjusting each annual APC to 2025 dollars using the CPI Advanced Economies, as reported by the International Monetary Fund (ibid.). For example, since the CPI Advanced Economies from 2019 to 2025 was 1.225, a listed 2019 APC of $3,000 was converted to $3,675 to obtain a 2025 equivalent value. Ultimately, since the academic journal market is an international endeavor with many parties across the globe involved in the production, no single price index fits it perfectly. We take the CPI Advanced Economies to be the most suitable indicator for a general inflation adjustment. Readers wishing to view comparisons using a different CPI are invited to do so using the data provided in Matthias et al. (2026b).

When reporting our results, we present both the inflation adjusted as well as unadjusted amounts. Inflation adjusted numbers are helpful to demonstrate trends in annual prices, while the unadjusted amounts represent actual fees paid, enabling libraries and funders to compare to their respective annual budgets. Figures and tables without inflation adjustment are provided in Supplement 2.

## 2.4 Calculating global APC spend

The number of APC-able documents per journal per year was multiplied by the annual APC data. This approach assumes every article was charged its journal's list price, and so does not account for discounts, waivers, or transformative agreements. Some journals switched publishers during the study period (n = 195). In these cases, the dataset retains a single APC record per journal-year, corresponding to the new publisher from the year of transfer onward (see Matthias et al., 2026b). We matched each journal-year's documents to that APC value. Finally, summary statistics such as sum, median, minimum and maximum were calculated for various aggregates such as by publisher, year, or OA status, with results presented in Section 3. All analysis code—including the SQL queries used to collect article counts and the R scripts used to compute spend estimates and produce the figures and tables in this paper—is openly available on GitHub (Schares & Matthias, 2026).

# 3. Results

We estimate that globally, the amount of APCs paid to the 14 publishers included in the Matthias et al. (2026b) dataset was $15.08 billion over the seven-year period analyzed, or $16.22 billion adjusted for inflation to 2025 prices (Table 1). Using inflation-adjusted estimates, the spend on OA fees quadrupled from $927.31 million in 2019 to $3.74 billion in 2025. Below we report results comparing estimated spend per OA type, publisher, and listed versus paid APCs. Corresponding publication counts are provided in Supplement 3.

**Table 1.** Estimate of annual APC spend (in millions USD) per OA type for actual APCs paid and adjusted for inflation using CPI Advanced Economies.

| Publication year | Spend estimate based on APCs in millions USD (actual list price) | | Spend estimate based on APCs in millions USD (adjusted to 2025 prices) | |
|---|---|---|---|---|
| | Gold | Hybrid | Gold | Hybrid |
| 2019 | 558.26 | 198.72 | 683.87 | 243.44 |
| 2020 | 788.59 | 340.94 | 959.72 | 414.92 |
| 2021 | 1,198.32 | 500.24 | 1,414.01 | 590.29 |
| 2022 | 1,520.84 | 638.00 | 1,672.93 | 701.80 |
| 2023 | 1,640.15 | 900.82 | 1,725.43 | 947.66 |
| 2024 | 1,921.14 | 1,136.17 | 1,969.16 | 1,164.57 |
| 2025 | 2,393.63 | 1,342.63 | 2,393.63 | 1,342.63 |
| 2019-2025 | 10,020.93 | 5,057.52 | 10,818.76 | 5,405.31 |

## 3.1 OA type

Using inflation-adjusted estimates, 66.7% of the spend went to gold OA. This reflects both the presence of high-volume gold-only publishers, such as MDPI and Frontiers, and the large gold portfolios of publishers that offer both models. Gold OA spend increased by 250.0% from $683.87 million in 2019 to $2,393.63 million in 2025, and hybrid OA spend by 451.5% from $243.44 to $1,342.63 million (Figure 2, Table 1).

**Estimate of annual APC spend (in USD) by OA type**

Adjusted for inflation to 2025 USD using CPI Advanced Economies.

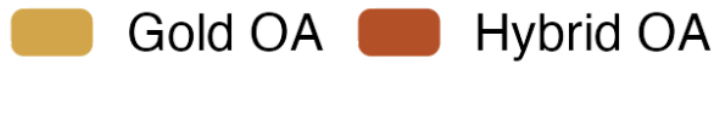

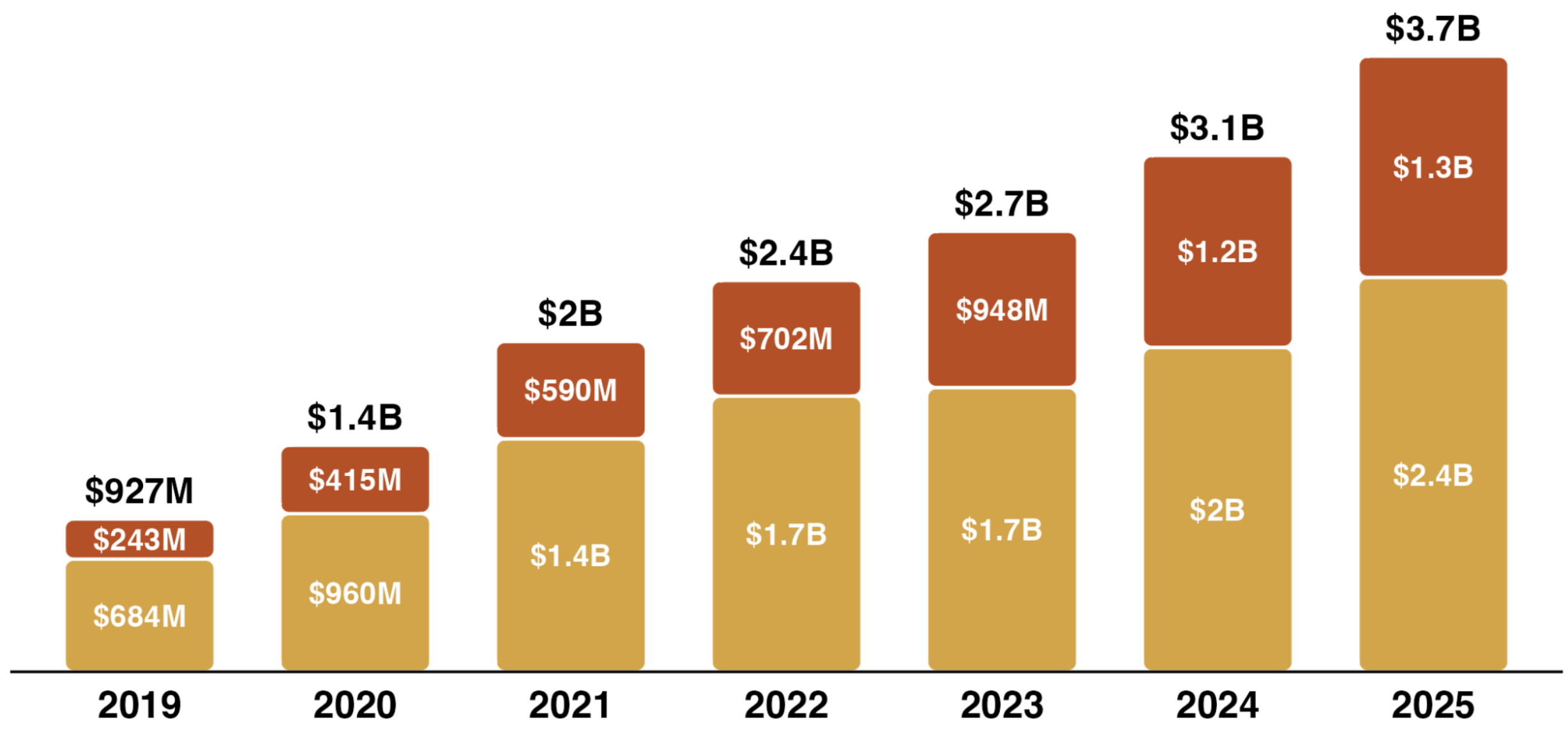

Some publishers are not covered in every year (De Gruyter and OUP begin in 2021, IOP is missing 2020, and Sage's 2019 and 2025 data are gold-only and hybrid-only respectively).

**Figure 2.** Estimate of annual APC spend (in USD) per OA type, adjusted for inflation to 2025 USD using CPI Advanced Economies.

## 3.2 Publishers

Estimated APC spend rose across nearly all publishers over the period, though individual trajectories differed. MDPI became the largest publisher by estimated APC revenue when it overtook Springer Nature in 2021 (Figure 3), though its spend then plateaued around $620–630 million from 2022 to 2024 even as its published output fell 17.4% over the same period—implying that rising per-article APCs offset the decline in volume to keep total revenue stable. Elsevier grew far faster than Springer Nature, overtaking it in 2023 and MDPI by 2024 to become the largest, with Springer Nature second and MDPI third by 2025. Wiley grew steadily throughout the period, reaching $494 million in 2025. Frontiers is the only publisher to show a sustained decline in annual APC revenue, dropping 25.4% from 2022 ($297.8 million) to 2023 ($222.3 million) and falling further through 2024 ($181.7 million) before a partial recovery in 2025. Frontiers' revenue decline and MDPI's plateau both coincide with falling publication output (Frontiers -29.7% from 2022 to 2023), which may be linked to scandals around paper mills (Matusz et al., 2025; Richardson et al., 2025), criticism of special issues as a growth model (Crosetto et al., 2026; Gleasner & Sood, 2025), and the delisting of journals from the Web of Science in March 2023 due to "increasing threats to the integrity of the scholarly record"

(Brainard, 2023; Brundy & Thornton, 2024; King, 2023). Related scandals and thousands of retractions led Wiley to shut down 19 journals from its Hindawi imprint, compromised by paper mills, and to retire the Hindawi name (Kincaid, 2023; Subbaraman, 2024). Despite these closures, our estimates show no corresponding disruption to Wiley's gold OA output or revenue.

Among the smaller publishers, most grew steadily over the period, with OUP nearly doubling its estimated revenue since 2021 to reach $113 million in 2025—a rise that coincided with its number of active ESAC-registered transformative agreements growing from 17 to 34 over the same period (ESAC, 2026) —while IEEE grew least over the seven years studied ($37.8 to $47.1 million, +24.6%). Sage’s apparent drop in 2025 reflects incomplete data coverage for that year rather than a verified decline.

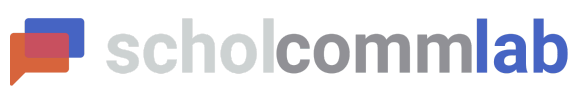

## Estimate of annual APC spend (in USD) by publisher

Adjusted for inflation to 2025 USD using CPI Advanced Economies.

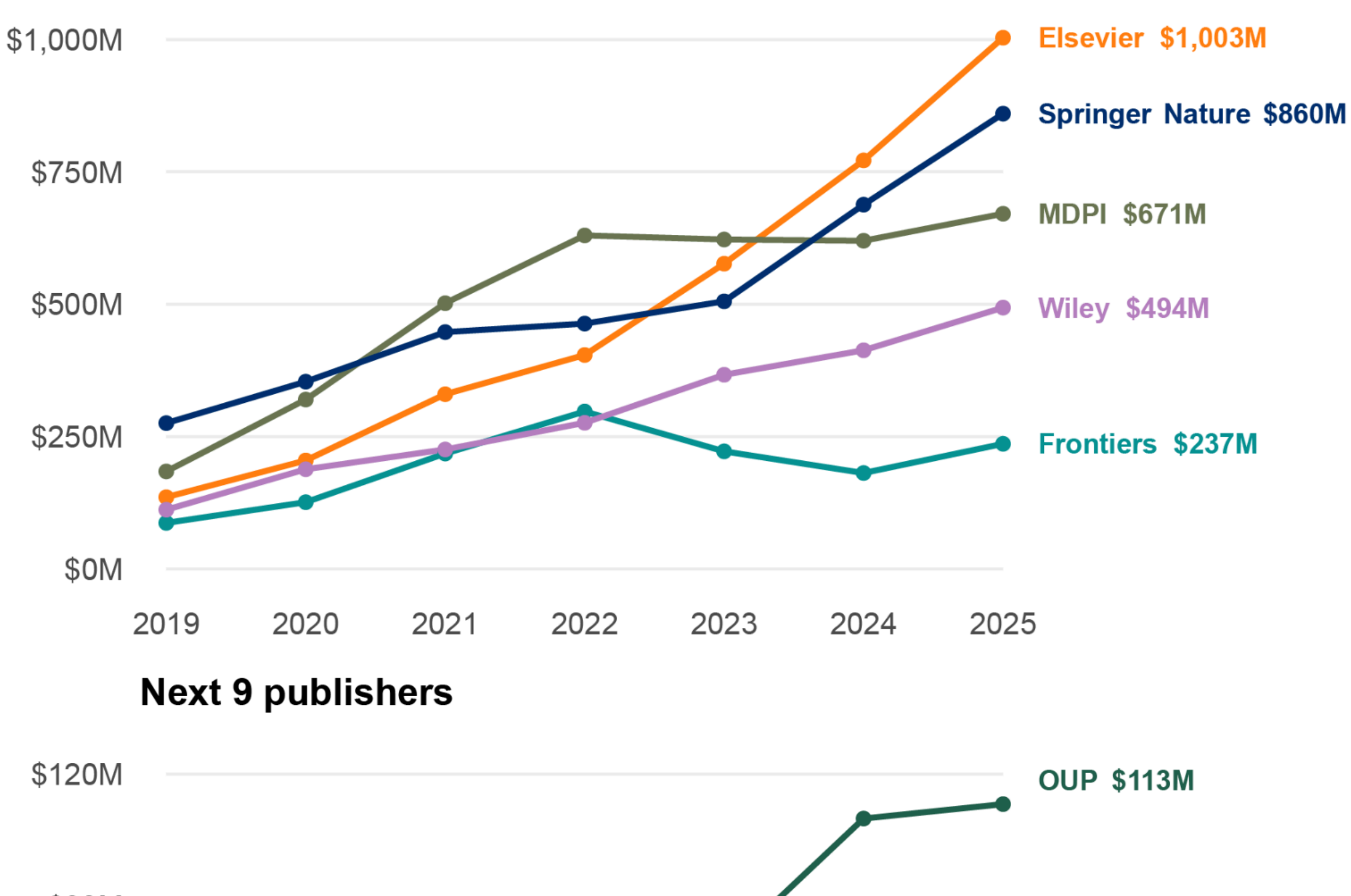

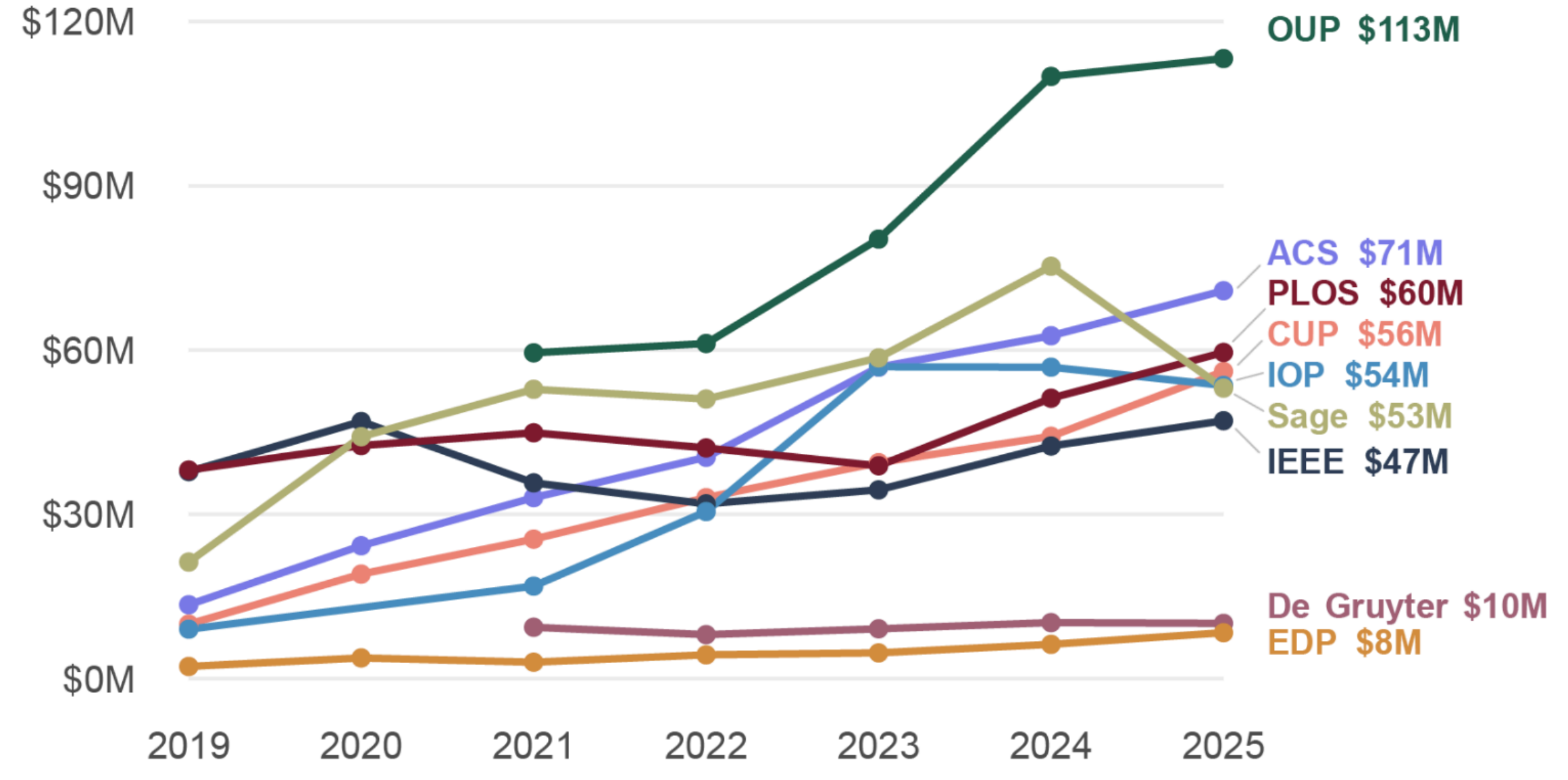

Sage's 2019 and 2025 figures reflect gold-only and hybrid-only data respectively.

**Figure 3.** Estimate of annual APC spend (in USD) by publisher adjusted for inflation to 2025 USD using CPI Advanced Economies.

Figure 4 shows estimated annual spend per publisher by OA type. Among publishers offering both routes, the balance between gold and hybrid varies considerably. Elsevier and Wiley draw more revenue from hybrid than gold, with Elsevier reaching $553 million in hybrid against $450 million in gold by 2025, and Wiley $281 million against $213 million. Springer Nature shows the opposite, with gold ($575 million) roughly double its hybrid revenue ($285 million) in 2025. The contrast is sharper at the smaller publishers: ACS brings in $58 million from hybrid against just $13 million from gold (4.5x), while IEEE is almost the mirror image, at $38 million gold against $9 million hybrid (4.2x). Several publishers changed course during the period. CUP's gold revenue overtook its hybrid revenue in 2025, and EDP's hybrid revenue collapsed after 2021 to $420,000, leaving it almost entirely gold. IOP shows a similar reversal, with hybrid peaking in 2022 before falling below gold.

**Estimate of annual APC spend (in USD) by publisher and OA type**

Adjusted for inflation to 2025 USD using CPI Advanced Economies.

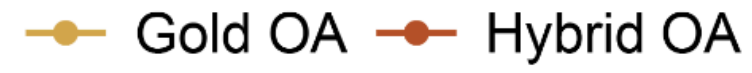

Elsevier
$600M
$400M
$200M
$0
$553M
$450M

Springer Nature
$600M
$400M
$200M
$0
$575M
$285M

MDPI
$600M
$400M
$200M
$0
$671M

Wiley
$300M
$200M
$100M
$0
$281M
$213M

Frontiers
$300M
$200M
$100M
$0
$237M

OUP
$60M
$40M
$20M
$0
$60M
$54M

ACS
$60M
$40M
$20M
$0
$58M
$13M

PLOS
$60M
$40M
$20M
$0
$60M

CUP
$60M
$40M
$20M
$0
$33M
$23M

IOP
$60M
$40M
$20M
$0
$32M
$22M

Sage
$60M
$40M
$20M
$0
$53M
$29M

IEEE
$60M
$40M
$20M
$0
$38M
$9M
2019 2022 2025

De Gruyter
$8M
$4M
$0
$6M
$4M
2019 2022 2025

EDP
$8M
$4M
$0
$8M
$420K
2019 2022 2025

Panels are ordered by total 2025 APC spend. Note that the y-axis scale differs between panels.

**Figure 4.** Estimate of annual APC revenue (in USD) by publisher and OA type adjusted for inflation to 2025 USD using CPI Advanced Economies.

Analyzing growth in APC spend by publisher (Table 2), Elsevier (638.4%), IOP (595.9%), and CUP (465.4%) showed the largest proportional increases in APC revenue from 2019 to 2025—each more than a fivefold rise, even after accounting for inflation. Looking at all publishers combined, spend grew faster than article output (+302.9% versus +237.6%), indicating that rising per-article fees, not volume growth alone, drove the increase in total spend. For Elsevier, EDP, and IEEE, however, spend and output grew at similar rates.

Spend on hybrid APCs generally grew faster than spend on gold (+451.5% versus +250.0%), driven by faster growth in hybrid article output (+427.4% versus +196.5%). This may be influenced by transformative agreements (Jahn, 2025a). At the publisher level the picture is more mixed: hybrid outgrew gold at four of the eight publishers offering both routes, including the two largest, Elsevier and Springer Nature. The divergence is sharpest at IEEE, where hybrid spend rose 175.5% while gold grew just 10.2%. Gold grew faster at the other four, most strikingly at CUP (+1,205.6% against 212.6%) and EDP (+1,130.8% against −72.9%), where gold increased thirteen- and twelvefold respectively.

**Table 2.** Growth of article output and APC spend (adjusted for inflation to 2025 USD using CPI Advanced Economies) from 2019 to 2025 per publisher and OA type.

| 2019 to 2025 growth rate | Number of APC-able publications | | | Spend estimate based on APCs in USD (adjusted to 2025 prices) | | |
|---|---|---|---|---|---|---|
| | Gold+Hybrid | Gold | Hybrid | Gold+Hybrid | Gold | Hybrid |
| **all publishers** | **+237.6%** | **+196.5%** | **+427.4%** | **+302.9%** | **+250.0%** | **+451.5%** |
| ACS | +326.9% | +133.9% | +629.6% | +426.2% | +383.2% | +436.6% |
| CUP | +372.4% | +648.3% | +199.1% | +465.4% | +1,205.6% | +212.6% |
| De Gruyter | n/a | n/a | n/a | n/a | n/a | n/a |
| EDP | +284.8% | +701.9% | -73.1% | +282.2% | +1,130.8% | -72.9% |
| Elsevier | +627.7% | +651.2% | +596.0% | +638.4% | +596.3% | +676.5% |
| Frontiers | +161.5% | +161.5% | n/a | +171.3% | +171.3% | n/a |
| IEEE | +27.2% | +15.8% | +176.9% | +24.6% | +10.2% | +175.5% |
| IOP | +422.9% | +475.3% | +357.2% | +490.9% | +595.9% | +385.5% |
| MDPI | +145.2% | +145.2% | n/a | +263.3% | +263.3% | n/a |
| OUP | n/a | n/a | n/a | n/a | n/a | n/a |
| PLOS | +30.6% | +30.6% | n/a | +56.4% | +56.4% | n/a |
| SAGE | n/a | n/a | n/a | n/a | n/a | n/a |
| Springer Nature | +172.3% | +150.6% | +255.7% | +212.3% | +188.3% | +275.1% |
| Wiley | +324.4% | +338.1% | +310.9% | +339.5% | +386% | +309.8% |

Note: n/a indicates that a growth rate could not be calculated. For De Gruyter, OUP, and Sage this reflects incomplete coverage—De Gruyter and OUP first appear in the dataset in 2021, and Sage's 2019 and 2025 records are gold-only and hybrid-only respectively. For Frontiers, MDPI, and PLOS, the absence of hybrid figures reflects their gold-only publishing model rather than a gap in the data.

## 3.3 APCs listed vs paid

Underlying these totals is considerable variation in APC price, and some price points attract far more articles than others. The violin plots in Figure 5 show the distribution of listed APC prices, where each price point represents one journal (left half, purple), and APCs weighted by the number of articles each journal published in 2025 (right half, orange). The right half is an estimate of what was paid.

We report medians throughout this section, since the distributions are skewed. Consistent with other studies (Asai, 2022; Björk & Solomon, 2014; Schönfelder, 2020, 2026), we find the median APCs offered by gold journals are lower ($2,160) than hybrid ($3,550). The median weighted, however, exceeds the median listed for both gold ($2,896) and hybrid ($3,810), indicating that authors publish disproportionately in journals with higher fees, or alternatively, that publishers charge higher fees for popular journals

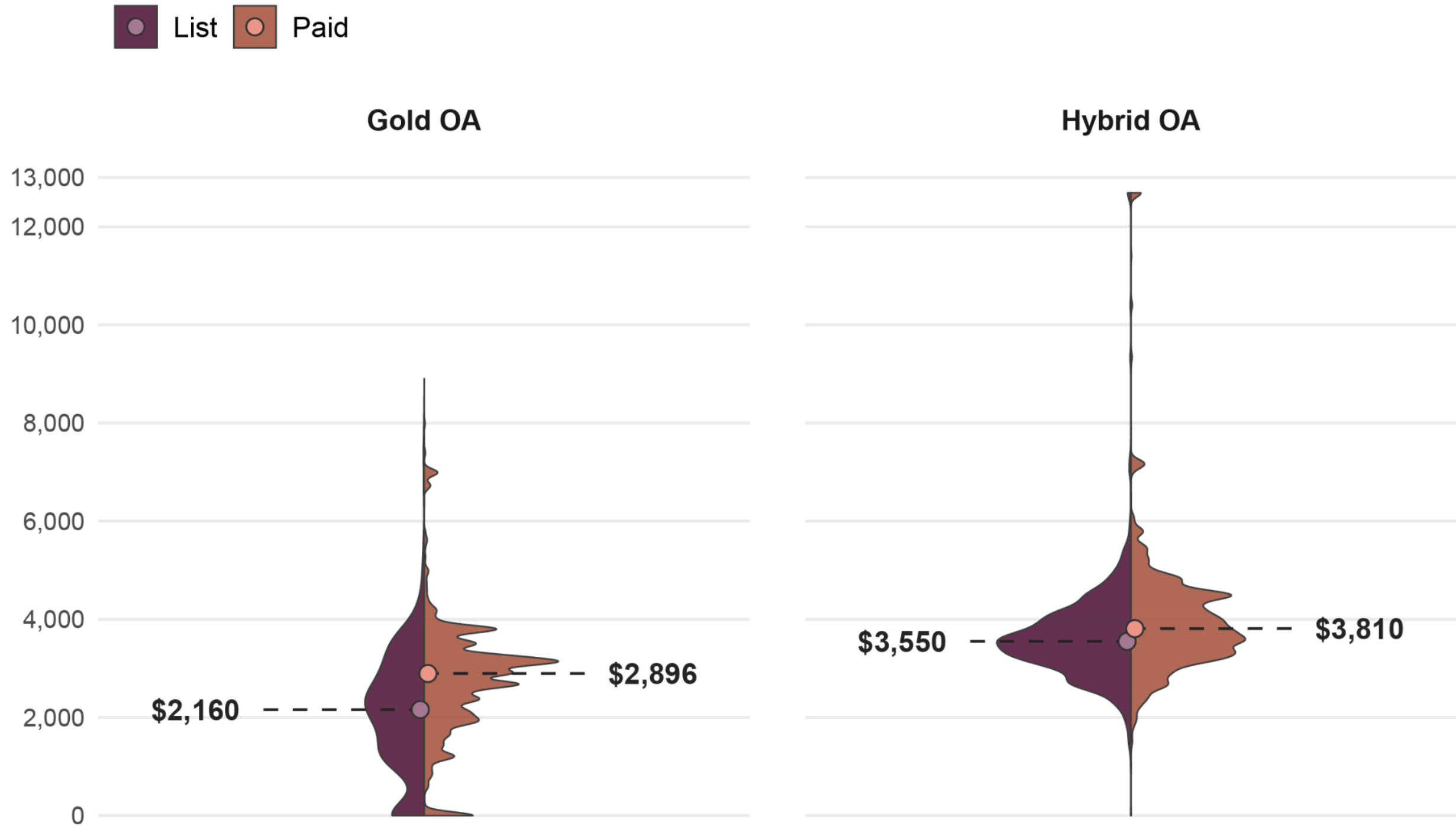

**Figure 5.** Distribution of the number of journals per listed APC (journal level) and the number of articles per paid APC (article level) for 2025 by OA type, all publishers. Medians are indicated with a dashed line.

Comparing across publishers (Table 3), median listed gold APCs range from $0 at De Gruyter and EDP—both of which operate substantial no-fee tiers within their gold portfolios—to $3,550 at CUP and $3,500 at ACS. Median listed hybrid APCs are consistently higher, from $1,870 at EDP to $4,500 at ACS. Weighted by article volume, median gold APCs rise at nine of the thirteen publishers with gold data for 2025, most sharply at MDPI ($1,450 to $3,140) and Frontiers ($2,350 to $3,800). At two publishers the pattern reverses: ACS ($3,500 to $1,940) and PLOS ($3,040 to $2,380) show lower paid than listed medians, indicating that their output concentrates in lower-priced journals. CUP and IEEE remain the same.

Figures 6 and 7 show that such differences are not evenly distributed across the market. Among the five publishers with the highest APC revenue in 2025—Elsevier, Springer Nature, MDPI, Wiley, and Frontiers—the median article costs more than the median journal in every case, for both gold and hybrid portfolios (Figure 6). At each of these publishers, more than half of all articles appeared in journals priced above the publisher's median journal. Among the nine lower-revenue publishers, no such pattern holds (Figure 7). Article volume skews higher than the listed prices of the gold portfolios of EDP, De Gruyter, and IOP, and of both portfolios of OUP. Elsewhere it tracks the list closely—at CUP and IEEE across both portfolios, and in the hybrid portfolios of ACS, De Gruyter, EDP, IOP, and Sage, where price lists offer little variation. The distribution of paid prices only skews lower for the gold portfolios of PLOS and ACS.

**Table 3.** Median of 2025 APCs listed (journal level) and paid (article level) per OA status per publisher.

| | Median gold APC 2025 (in USD) | | Median hybrid APC 2025 (in USD) | |
|---|---|---|---|---|
| | Listed | Paid | Listed | Paid |
| **all publishers** | **2,160** | **2,900** | **3,550** | **3,810** |
| ACS | 3,500 | 1,940 | 4,500 | 4,500 |
| CUP | 3,550 | 3,550 | 3,550 | 3,550 |
| De Gruyter | 0.00 | 848.00 | 2,710 | 2,710 |
| EDP | 0.00 | 1,870 | 1,870 | 1,870 |
| Elsevier | 1,890 | 2,120 | 3,530 | 3,770 |
| Frontiers | 2,350 | 3,800 | n/a | n/a |
| IEEE | 2,080 | 2,080 | 2,650 | 2,650 |
| IOP | 2,600 | 3,010 | 3,330 | 3,330 |
| MDPI | 1,450 | 3,140 | n/a | n/a |
| OUP | 2,600 | 2,960 | 4,110 | 4,330 |
| PLOS | 3,040 | 2,380 | n/a | n/a |
| Sage | n/a | n/a | 3,650 | 3,650 |
| Springer Nature | 2,490 | 2,790 | 3,290 | 3,490 |
| Wiley | 2,400 | 2,910 | 3,730 | 4,150 |

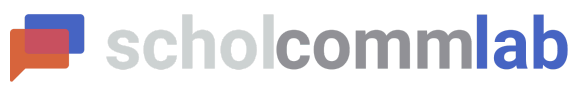

## Listed and paid APCs (in USD), 2025: Top 5 publishers

In every case the median article costs more than the median journal: most articles appear in the more expensive part of the portfolio.

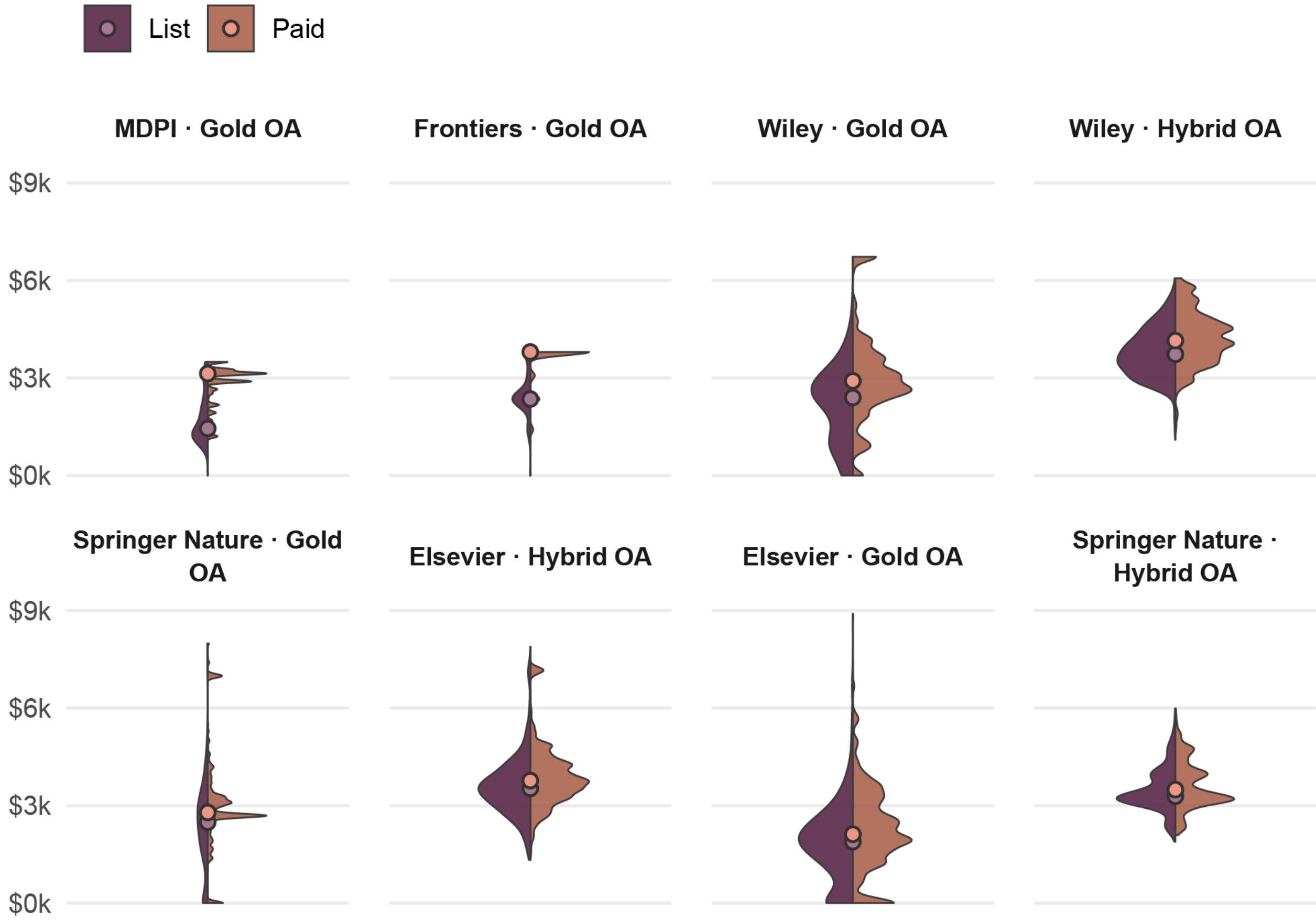

Panels ordered by size of the gap. Left half: listed prices, one per journal. Right half: weighted by article volume. Points mark medians. Journals above $9,000 not shown: 20 at Elsevier (1.1% of hybrid titles, 0.8% of hybrid article volume) and 39 at Springer Nature (1.9% of hybrid titles, 4.8% of hybrid volume).

**Figure 6.** Distribution of listed (journal-level, blue) and paid (article-level, orange) APCs for the top five publishers by OA type, 2025.

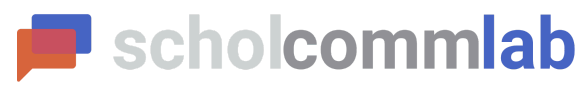

## Listed and paid APCs (in USD), 2025: Next 9 publishers

Article volume shifts up the price list at some publishers, tracks it at others, and shifts down at the rest.

List Paid

EDP · Gold OA
De Gruyter · Gold OA
IOP · Gold OA
OUP · Gold OA
OUP · Hybrid OA
ACS · Hybrid OA
CUP · Gold OA
CUP · Hybrid OA
De Gruyter · Hybrid OA
EDP · Hybrid OA
IEEE · Gold OA
IEEE · Hybrid OA
IOP · Hybrid OA
Sage · Hybrid OA
PLOS · Gold OA
ACS · Gold OA

$6k
$3k
$0k

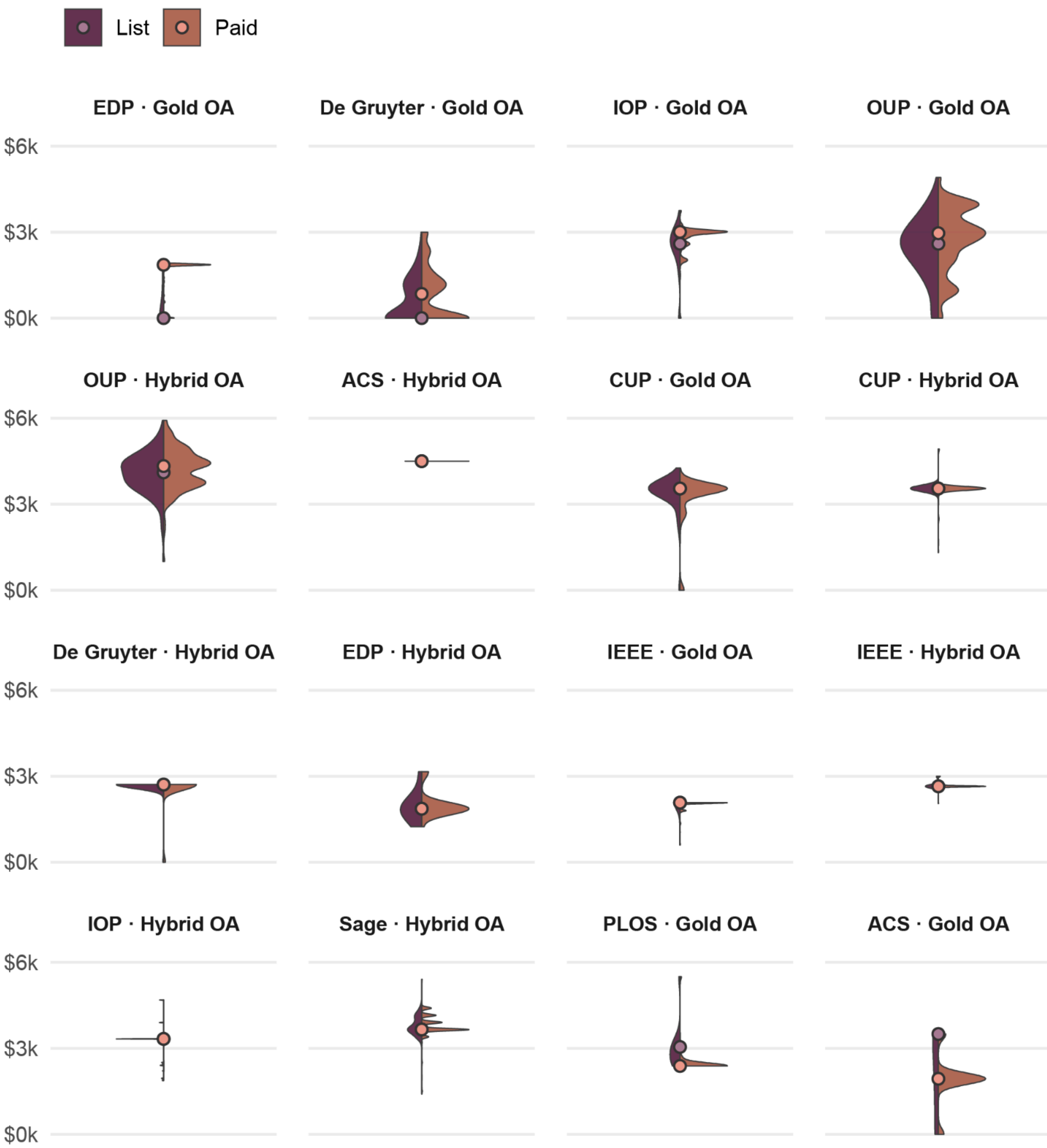

Panels grouped by direction of shift, then ordered by size of the gap. Left half: listed prices, one per journal. Right half: weighted by article volume. Points mark medians. Four journals above $6,000 not shown (3 hybrid at OU P 1 gold at PLOS; under 1% of article volume at each).

**Figure 7.** Distribution of listed (journal-level, blue) and paid (article-level, orange) APCs for the remaining nine publishers by OA type, 2025.

# 4. Discussion

We estimate that $3.7 billion was spent on APCs across the 14 publishers in our sample in 2025, of which 87% went to five publishers—Elsevier, Springer Nature, MDPI, Wiley, and Frontiers. This concentration has held steady throughout the period, with these top five accounting for 85–87% of total spend in every year since 2019. Global spend has quadrupled in seven years, even after adjusting for inflation, driven by a growing volume of gold and hybrid articles and rising fees. Elsevier accounts for the largest share of that growth. We estimate that their APC revenue grew 638% between 2019 (adjusted) and 2025, more than double the 303% across all publishers, taking Elsevier from third position to first and past $1 billion annually. Several smaller publishers also grew faster than the market average—CUP at 465% and IOP at 491%—but from bases small enough that their gains remain a fraction of market share: CUP added roughly $46 million in annual revenue over the period, against Elsevier's $867 million.

Gold OA still accounts for the larger share of APC expenditure—$2.4 billion of the $3.7 billion we estimate for 2025—but hybrid spending grew fastest. Hybrid revenue rose 451% between 2019 and 2025 against 250% for gold, driven by a corresponding surge in hybrid article output (+427% against +197%). This runs counter to the direction of OA policy over the same period, which has focused on transitioning away from the hybrid model rather than expanding it. Transformative agreements (TAs)—introduced as a temporary mechanism to shift academic publishing from subscriptions toward full OA and increasingly adopted in recent years (Campbell et al., 2022; Steinberg, 2025)—are a potential contributing factor to this trend as previous studies indicate that TAs increase hybrid OA publishing (Bakker et al., 2024; Sterman et al., 2025). Publishers report similar effects: Springer Nature states that its TAs produced ten times more OA articles in 2024 than author choice alone, up from seven times in 2023 (Springer Nature, 2025).

These agreements may also help explain the gap between median list prices and the median APCs for the articles published. TAs typically include APCs for entire journal portfolios, but sometimes only for hybrid journals; for instance, the Canadian Research Knowledge Network's agreement with Elsevier covers fees for over 1,800 hybrid journals, and its agreement with CUP covers APCs across both gold and hybrid (CRKN, n.d.). Authors supported by such agreements would be less price sensitive than those paying APCs from their research grants, perhaps explaining part of the pattern we observe at the largest publishers—article volume concentrating in journals priced above the publisher's median. Another likely explanation why higher-priced journals attract disproportionate output in the first place is prestige. If APCs track a journal's standing rather than its production costs, this would help explain the disconnect between cost and APC price (Grossmann & Brembs, 2021; Khoo, 2019).

Interpreting these price patterns depends on how carefully "price" itself is measured over time—which is where our approach departs from prior work. Our method improves on prior estimates by accounting for price changes year by year, rather than applying current rates to historical publication volumes. This matters because studies using a single current APC rate across years can substantially overestimate past spending (Kendall, 2024). By linking journal-level prices to the number of APC-able articles published each year, our estimates avoid this bias and can support more precise analysis of institutional or national spending over time. Where methodological choices had to be made—for example, in handling non-flat-rate fee types (Section 2.1)—we documented our assumptions, so that readers can assess and, if needed, recompute estimates under different choices. We also worked to avoid inflating our count of APC-able articles, filtering out additional non-article document types before matching against price data (see Supplement 1).

Nevertheless, our approach has several limitations. First, due to the lack of transparent information about the amount of OA fees paid, we are *estimating* global spend based on APC list prices and published output; only publishers know what they in fact receive, and we encourage them to report it. Second, our dataset does not include Taylor & Francis, one of the largest commercial publishers (Butler et al., 2023). Unlike the other publishers in our sample, Taylor & Francis does not publish a static, downloadable APC price list; APCs are instead retrieved through an interactive online tool that returns prices only for journals queried individually, is not amenable to bulk collection or Wayback Machine archiving, and shows prices that vary by the payer's country or region. As a result, our estimates of global APC spend and market concentration exclude one of the largest commercial publishers.

Third, beyond those identified in list prices or through manual checks, we cannot account for waivers and discounts. Of the 69,400 journal-year combinations in our dataset with a known APC, 1,989 were listed as $0, which may reflect a permanent waiver, society or third-party funding, or a temporary promotional waiver. However, waivers granted to individual authors are not captured in public metadata, and we cannot account for them. Several publishers offer waiver or discount programmes, but most do not publish data granular or regularly enough to include in our estimates (Frontiers, n.d.; IOP Publishing, 2023; McKenna, 2025; PLOS, n.d.; A. Smith et al., n.d.). Overall, we acknowledge that waivers were likely granted to some fraction of gold OA and hybrid OA articles, but we lack data to say more. Finally, we do not account for TAs (Borrego et al., 2020; ESAC, n.d.; Farley et al., 2021), so our calculations might therefore overestimate *fees paid by individual authors*. However, because such agreements are typically priced on publication volume or based on existing subscription costs, they amount to *fees paid up-front by libraries*. As such, we argue that our estimate of global spend still reflects the total amount of OA *fees paid by the community*.

# 5. Conclusion and Future Work

Future studies could extend this work in several directions. Triangulating publication counts across OpenAlex, Dimensions, and Web of Science would further refine the article counts underlying our spend estimates. A less examined trend is the monetization of green OA itself, through mechanisms such as article development charges and repository licensing fees (ACS Newsroom, 2023; IEEE Open, 2025); as these spread, they will warrant inclusion in future estimates of OA spend. Finally, extending this analysis to author affiliations to understand who pays APCs, as well as to investigate the effect of national OA policies on who pays, would be a valuable next step.

Our study contributes to ongoing debates about the APC model—its inequities, its unsustainability, and its disconnect from the actual cost of OA publishing. These debates have largely proceeded without a comprehensive evidence base. Our contribution is a dataset and analysis that can help ground them, supporting informed, evidence-based decisions by libraries, consortia, funders, and researchers as they navigate where to publish and which OA mechanisms are most reasonable to pursue. Even accounting for the limitations of our estimates, the scale is clear: billions of dollars are now spent annually on APCs, and spending on them has grown faster than the volume of articles published, reinforcing the disconnect discussed above.

Producing a dataset with this coverage was resource-intensive, and extending it to further publishers or years will be costly. That such an effort is necessary at all reflects a failure of transparency on the part of publishers. We call on them to disclose their OA revenues, disaggregated by journal, and to provide the community with comprehensive data about the volume and sources of funds behind APCs, transformative agreements, rapid service fees, article development charges, and any other fees paid by academic societies, institutions or governments. Until publishers do, estimates like ours will remain the best available account of what the OA transition is costing the academic community.

## Data availability

The APC dataset analyzed in this paper is available on the Harvard Dataverse under a CC-0 license. It is cited as Matthias et al. (2026b) in the reference list and available at https://doi.org/10.7910/DVN/AZ985C.

Supporting code and data underlying this analysis, including the SQL query used to collect article counts and R scripts to compute spend estimates and produce figures and tables, are available at https://github.com/ScholCommLab/apc-analysis.

## Acknowledgements

We would like to thank Jason Portenoy and Kyle Demes at OpenAlex for resolving issues about OA status and document types. We also thank Jeffrey Brainard, Diego Kozlowski, Martin Reinhart and Ted Bergstrom for fruitful discussions about inflation adjustments. We would also like to thank Najko Jahn and the SUB Göttingen team for providing access to the OpenAlex snapshot on GBQ.

## Author contributions

L.M.: Data curation, Formal analysis, Investigation, Methodology, Validation, Visualization, Writing—original draft, Writing—review & editing; E.S.: Conceptualization, Data curation, Formal analysis, Investigation, Methodology, Validation, Writing—original draft, Writing—review & editing; J-P.A.: Conceptualization, Data curation, Formal analysis, Investigation, Methodology, Writing—review & editing; L-A.B.: Conceptualization, Writing—original draft, Writing—review & editing; S.K.: Validation, Writing—original draft, Writing—review & editing; N.S.: Data curation; S.H.: Conceptualization, Data curation, Investigation, Methodology, Project administration, Resources, Supervision

## Competing interests

JPA and SH maintain a financial interest in pending litigation against four of the publishers discussed in this paper, which seeks recovery for allegedly unreasonable and unnecessary article processing charges.

## Funding information

This research was funded by the Volkswagen Foundation (Az.: 9C784).

# References

ACS Newsroom. (2023, September 21). *ACS Publications provides a new option to support zero-embargo green open access*. https://www.acs.org/pressroom/newsreleases/2023/september/acs-publications-provides-new-option-to-support-zero-embargo-green-open-access.html

Akbulut, M. (2026). The Transformative Agreement paradox: A game theory analysis of large scholarly publishers and library negotiation strategies. *Scientometrics*, *131*(1), 647–664. https://doi.org/10.1007/s11192-025-05496-8

Alperin, J. P., Portenoy, J., Demes, K., Larivière, V., & Haustein, S. (2024). *An analysis of the suitability of OpenAlex for bibliometric analyses* (Version 1). arXiv. https://doi.org/10.48550/ARXIV.2404.17663

Asai, S. (2022). Determinants of article processing charges for hybrid and gold open access journals. *Information Discovery and Delivery*, *51*(2), 121–129. https://doi.org/10.1108/IDD-09-2021-0098

Asai, S. (2023). Which database with article processing charges should be used? *Scientometrics*, *128*(11), 6293–6298. https://doi.org/10.1007/s11192-023-04841-z

Asai, S. (2024). Determinants of manuscript submissions to fully open access journals: Elasticity to article processing charges. *Scientometrics*, *129*(3), 1687–1696. https://doi.org/10.1007/s11192-024-04934-3

Ayeni, P., & Larivière, V. (2025). Inequity, precarity, and disparity: Exploring systemic and institutional barriers in open access publishing. *Journal of Librarianship and Information Science*. https://doi.org/10.1177/09610006251353385

Bakker, C., Langham-Putrow, A., & Riegelman, A. (2024). The Impact of Transformative Agreements on Publication Patterns: An Analysis Based on Agreements from the ESAC Registry. *International Journal of Librarianship*, *8*(4), 67–96. https://doi.org/10.23974/ijol.2024.vol8.4.341

Baldwin, J., & Cavanagh, P. (2024). When will we be transformed? Reflections on the experience of working with transformative agreements as a cross-library working group. *Insights*. https://doi.org/10.1629/uksg.645

Björk, B.-C., & Solomon, D. (2014). *Developing an Effective Market for Open Access Article Processing Charges* (Final report). https://cms.wellcome.org/sites/default/files/developing-effective-market-for-open-access-article-processing-charges-mar14.pdf

Borrego, Á. (2023). Article processing charges for open access journal publishing: A review. *Learned Publishing*, *36*(3), 359–378. https://doi.org/10.1002/leap.1558

Borrego, Á., Anglada, L., & Abadal, E. (2020). Transformative agreements: Do they pave the way to open access? *Learned Publishing*, *34*(2), 216–232. https://doi.org/10.1002/leap.1347

Brainard, J. (2023). Fast-growing open-access journals stripped of coveted impact factors. *Science*, *379*(6639), 1283–1284. https://doi.org/https://doi.org/10.1126/science.adi0098

Brundy, C., & Thornton, J. B. (2024). The paper mill crisis is a five-alarm fire for science: What can librarians do about it? *Insights*, *37*(1). https://doi.org/10.1629/uksg.659

Butler, L.-A. (2023). *Funding the Business of Open Access: A Bibliometric Analysis of Article Processing Charges, Research Funding, and the Revenues of the Oligopoly of Publishers* [Université d'Ottawa / University of Ottawa]. https://ruor.uottawa.ca/items/a4ccb284-f998-4d80-8144-2d6d64369d9d

Butler, L.-A., Hare, M., Schönfelder, N., Schares, E., Alperin, J. P., & Haustein, S. (2024a). *An open dataset of article processing charges from six large scholarly publishers (2019-2023)* (arXiv:2406.08356). arXiv. https://doi.org/10.48550/arXiv.2406.08356

Butler, L.-A., Hare, M., Schönfelder, N., Schares, E., Alperin, J. P., & Haustein, S. (2024b). *Open dataset of annual Article Processing Charges (APCs) of gold and hybrid journals published by Elsevier, Frontiers, MDPI, PLOS, Springer-Nature and Wiley 2019-2023* (Version 1) [Dataset]. Harvard Dataverse. https://doi.org/10.7910/DVN/CR1MMV

Butler, L.-A., Matthias, L., Simard, M.-A., Mongeon, P., & Haustein, S. (2023). The oligopoly's shift to open access: How the big five academic publishers profit from article processing charges. *Quantitative Science Studies*, *4*(4), 778–799. https://doi.org/10.1162/qss_a_00272

Campbell, C., Dér, Á., Geschuhn, K., & Valente, A. (2022). *How are transformative agreements transforming libraries?* 87th IFLA World Library and Information Congress, Dublin, Ireland. https://repository.ifla.org/rest/api/core/bitstreams/6f038d8f-004b-4c1b-8937-53d380b11847/content

Crosetto, P., Barreiro, P. G., & Hanson, M. A. (2026). *The Issue with Special Issues: When Guest Editors Publish in Support of Self* (arXiv:2601.07563). arXiv. https://doi.org/10.48550/arXiv.2601.07563

Culbert, J. H., Hobert, A., Jahn, N., Haupka, N., Schmidt, M., Donner, P., & Mayr, P. (2025). Reference coverage analysis of OpenAlex compared to Web of Science and Scopus. *Scientometrics*, *130*(4), 2475–2492. https://doi.org/10.1007/s11192-025-05293-3

Cumming, D. J. (2026). *The Revenue of Finance Journals: Networks, Pricing Power, and Publication Volume* (SSRN Scholarly Paper No. 6885398). Social Science Research Network. https://papers.ssrn.com/abstract=6885398

de Jonge, H., Kramer, B., & Sondervan, J. (2025). Tracking transformative agreements through open metadata: Method and validation using Dutch Research Council NWO funded papers. *Quantitative Science Studies*, *6*, 1215–1227. https://doi.org/10.1162/QSS.a.24

ESAC. (n.d.). *Transformative Agreements—What are transformative agreements?* ESAC Initiative. Retrieved July 29, 2026, from https://esac-initiative.org/about/transformative-agreements/

ESAC. (2026). *ESAC Registry of Open Publishing Agreements* [Dataset]. https://esac-initiative.org/about/transformative-agreements/agreement-registry/

Farley, A., Langham-Putrow, A., Shook, E., Sterman, L. B., & Wacha, M. (2021). Transformative agreements: Six myths, busted. *College & Research Libraries News*, *82*(7), 298. https://doi.org/10.5860/crln.82.7.298

Frontiers. (n.d.). *Fee policy*. Retrieved July 27, 2026, from https://www.frontiersin.org/about/fee-policy

Gallardo, O., Milia, M., Appel, A. L., Team, G.-A., & Schalkwyk, F. van. (2024). *When researchers pay to publish: Results from a survey on APCs in four countries* (arXiv:2410.12144). arXiv. https://doi.org/10.48550/arXiv.2410.12144

Gleasner, R. M., & Sood, A. (2025). Special issues: The roles of special issues in scholarly communication in a changing publishing landscape. *Learned Publishing*, *38*(1), e1635. https://doi.org/10.1002/leap.1635

Grossmann, A., & Brembs, B. (2021). Current market rates for scholarly publishing services. *F1000Research*. https://doi.org/10.12688/f1000research.27468.2

Halevi, G., & Walsh, S. (2021). Faculty Attitudes Towards Article Processing Charges for Open Access Articles. *Publishing Research Quarterly*, *37*(3), 384–398. https://doi.org/10.1007/s12109-021-09820-x

Haustein, S., Schares, E., Alperin, J. P., Hare, M., Butler, L.-A., & Schönfelder, N. (2024). *Estimating global article processing charges paid to six publishers for open access between 2019 and 2023* (arXiv:2407.16551). arXiv. https://doi.org/10.48550/arXiv.2407.16551

IEEE Open. (2025, March 14). *Repository License Fee*. https://open.ieee.org/repository-license-fee/

International Monetary Fund. (2026). *World Economic Outlook (April 2026)—Inflation rate, average consumer prices* [Dataset]. https://www.imf.org/external/datamapper/PCPIPCH@WEO

IOP Publishing. (2023). *Get published and cited*. https://publishingsupport.iopscience.iop.org/wp-content/uploads/2023/03/Author-Strategy-Guide-02.2023.pdf

Jahn, N. (2025a). Estimating transformative agreement impact on hybrid open access: A comparative large-scale study using Scopus, Web of Science and open metadata. *Scientometrics*, *131*, 901–924. https://doi.org/10.1007/s11192-025-05390-3

Jahn, N. (2025b). How open are hybrid journals included in transformative agreements? *Quantitative Science Studies*, *6*, 242–262. https://doi.org/10.1162/qss_a_00348

Jahn, N., & Haupka, N. (2022, June 7). How open are hybrid journals included in nationwide transformative agreements in Germany? *Scholarly Communication Analytics*. https://subugoe.github.io/scholcomm_analytics/posts/oam_hybrid/

Jahn, N., Hobert, A., & Haupka, N. (2021). Entwicklung und Typologie des Datendiensts Unpaywall. *Bibliothek Forschung und Praxis*, *45*(2), 293–303. https://doi.org/10.1515/bfp-2020-0115

Jahn, N., Matthias, L., & Laakso, M. (2022). Toward transparency of hybrid open access through publisher-provided metadata: An article-level study of Elsevier. *Journal of the Association for Information Science and Technology*, *73*(1), 104–118. https://doi.org/10.1002/asi.24549

Jung, Y., Lee, J. Y., Lee, J., An, B.-G., Kim, W. J., & Park, J. (2025). Article processing charge costs of open access articles indexed in the Web of Science Core Collection from 2019 to 2023 by publisher and country: A secondary publication. *Science Editing*, *12*(2), 114–123. https://doi.org/10.6087/kcse.370

Kemp, J., & Skinner, K. (2024). *The Cost and Price of Public Access to Scholarly Publications: A Synthesis* (Version v1). Invest in Open Infrastructure. Zenodo. https://doi.org/10.5281/zenodo.14013060

Kendall, G. (2024). Are open access fees a good use of taxpayers' money? *Quantitative Science Studies*, *5*(1), 264–270. https://doi.org/10.1162/qss_c_00305

Khoo, S. Y.-S. (2019). Article Processing Charge Hyperinflation and Price Insensitivity: An Open Access Sequel to the Serials Crisis. *LIBER Quarterly: The Journal of the Association of European Research Libraries*, *29*(1), 1–18. https://doi.org/10.18352/lq.10280

Kincaid, E. (2023, December 6). Wiley to stop using "Hindawi" name amid $18 million revenue decline. *Retraction Watch*. https://retractionwatch.com/2023/12/06/wiley-to-stop-using-hindawi-name-amid-18-million-revenue-decline/

King, A. (2023, April 24). Sanctioning of 50 journals raises concerns over special issues in 'mega-journals.' *Chemistry World*. https://www.chemistryworld.com/news/sanctioning-of-50-journals-raises-concerns-over-special-issues-in-mega-journals/4017315.article

Kulczycki, E., Alonso-Gamboa, J. O., Beigel, F., Digiampietri, L., Laakso, M., Pölönen, J., Taşkın, Z., & Cuartas, G. V. (2026). Beyond the Oligopoly: Scholarly Journal Publishing

Landscapes in Latin America and Europe. *Journal of Data and Information Science*. https://doi.org/10.1515/jdis-2025-0440

Matthias, L., Chavarro, D., Schares, E., Alperin, J. P., Rose, M., Frost, M., Camargo, F., Höfting, J., Butler, L.-A., Schönfelder, N., & Haustein, S. (2026a). *A dataset of article processing charges from 14 scholarly publishers, 2019–2025*. arXiv. https://doi.org/10.48550/arXiv.2608.14116

Matthias, L., Chavarro, D., Schares, E., Alperin, J. P., Rose, M., Frost, M., Camargo, F., Höfting, J., Butler, L.-A., Schönfelder, N., & Haustein, S. (2026b). *An open dataset of article processing charges from 14 scholarly publishers, 2019–2025* [Dataset]. Harvard Dataverse. https://doi.org/https://doi.org/10.7910/DVN/AZ985C

Matusz, P. J., Abalkina, A., & Bishop, D. V. M. (2025). The Threat of Paper Mills to Biomedical and Social Science Journals: The Case of the Tanu.pro Paper Mill in Mind, Brain, and Education. *Mind, Brain, and Education*, *19*(2), 90–100. https://doi.org/10.1111/mbe.12436

McKenna, J. (2025, October 13). How MDPI Informs the Academic Community About Open Access. *MDPI Blog*. https://mdpiblog.wordpress.sciforum.net/2025/10/13/open-access-information/

Neylon, C., & Kramer, B. (2026). *Trajectory modelling of publishing costs using Open Research Information—An open source analytics pipeline*. Zenodo. https://doi.org/10.5281/zenodo.20957444

Nicholas, D., Revez, J., Abrizah, A., Rodríguez-Bravo, B., Boukacem-Zeghmouri, C., Clark, D., Xu, J., Swigon, M., Watkinson, A., Jamali, H. R., & Herman, E. (2024). Purchase and publish: Early career researchers and open access publishing costs. *Learned Publishing*, *37*(4), e1617. https://doi.org/10.1002/leap.1617

OpenAPC. (2026). *OpenAPC* (Version 5.3.56-51-17) [Dataset]. GitHub. https://github.com/OpenAPC/openapc-de/tree/v5.3.56-51-17

ORION. (2026, February 16). *Open Research Information on BigQuery (ORION)*. ORION-DBs. https://orion-dbs.community/

Pieper, D., & Broschinski, C. (2018). OpenAPC: A contribution to a transparent and reproducible monitoring of fee-based open access publishing across institutions and nations. *Insights*, *31*(0). https://doi.org/10.1629/uksg.439

Piwowar, H., Priem, J., Larivière, V., Alperin, J. P., Matthias, L., Norlander, B., Farley, A., West, J., & Haustein, S. (2018). The state of OA: A large-scale analysis of the prevalence and impact of Open Access articles. *PeerJ*, *6*, e4375. https://doi.org/10.7717/peerj.4375

Piwowar, H., Priem, J., & Orr, R. (2019). *The Future of OA: A large-scale analysis projecting Open Access publication and readership* (p. 795310). bioRxiv. https://doi.org/10.1101/795310

PLOS. (n.d.). *Financial Overview—2024*. Retrieved July 27, 2026, from https://plos.org/financial-overview/

Priem, J., Piwowar, H., & Orr, R. (2022). *OpenAlex: A fully-open index of scholarly works, authors, venues, institutions, and concepts* (Version 2). arXiv. https://doi.org/10.48550/ARXIV.2205.01833

Richardson, R. A. K., Hong, S. S., Byrne, J. A., Stoeger, T., & Amaral, L. A. N. (2025). The entities enabling scientific fraud at scale are large, resilient, and growing rapidly.

*Proceedings of the National Academy of Sciences*, *122*(32), e2420092122. https://doi.org/10.1073/pnas.2420092122

Riegelman, A., & Langham-Putrow, A. (2025). Scoping Review of Transformative Agreement Research. *Evidence Based Library and Information Practice*, *20*(4), 252–352. https://doi.org/10.18438/eblip30721

Rodrigues, R. S., Abadal, E., & de Araújo, B. K. H. (2020). Open access publishers: The new players. *PloS One*, *15*(6), e0233432. https://doi.org/10.1371/journal.pone.0233432

Rothfritz, L., Schmal, W. B., & Herb, U. (2024). *Trapped in Transformative Agreements? A Multifaceted Analysis of >1,000 Contracts* (arXiv:2409.20224). arXiv. https://doi.org/10.48550/arXiv.2409.20224

Schares, E. (2023). Rethinking Hybrid Open Access Labels for Post-Embargo Articles. *OpenISU*. https://doi.org/10.31274/b8136f97.45a4f699

Schares, E., & Matthias, L. (2026). *ScholCommLab/apc-analysis* [Source code]. GitHub. https://github.com/ScholCommLab/apc-analysis

Schimmer, R., Geschuhn, K. K., & Vogler, A. (2015). *Disrupting the subscription journals' business model for the necessary large-scale transformation to open access*. https://doi.org/10.17617/1.3

Schmal, W. B. (2024a). How transformative are transformative agreements? Evidence from Germany across disciplines. *Scientometrics*, *129*(3), 1863–1889. https://doi.org/10.1007/s11192-024-04955-y

Schmal, W. B. (2024b). *The "Must Stock" Challenge in Academic Publishing: Pricing Implications of Transformative Agreements* (SSRN Scholarly Paper No. 4818955). Social Science Research Network. https://doi.org/10.2139/ssrn.4818955

Schönfelder, N. (2020). Article processing charges: Mirroring the citation impact or legacy of the subscription-based model? *Quantitative Science Studies*, *1*(1), 6–27. https://doi.org/10.1162/qss_a_00015

Schönfelder, N. (2026, March 17). *APC trends for gold and hybrid OA uncovered: Analysis of an unpublished, longitudinal dataset of list-price APCs for 7.000 hybrid journals collected by the oa.finder*. openCost: Navigating Cost Transparency, Hamburg, Germany. https://doi.org/10.3204/PUBDB-2026-00919

Schönfelder, N., & Tummes, J.-P. (2024). *Publikationsreport 2024: Ermittlung des Publikations-Outputs und der Open-Access-Anteile von Hochschulen des Landes Nordrhein-Westfalen*. https://doi.org/10.4119/unibi/2993325

Shu, F., & Larivière, V. (2024). The oligopoly of open access publishing. *Scientometrics*, *129*(1), 519–536. https://doi.org/10.1007/s11192-023-04876-2

Smith, A., Wheeler, A., Moore, L., Stimson, L., Giampoala, M., Lehnert-Bechle, M., Willis, M., McMullin, T., & Gaston, T. (n.d.). *Working Together to Create an Equitable Open Access System*. Wiley. Retrieved July 27, 2026, from https://www.wiley.com/en-ca/publish/article/open-access/insights/equitable-open-access-system/

Smith, M., Anderson, I., Bjork, B.-C., McCabe, M., Solomon, D., Tananbaum, G., Tenopir, C., & Willmott, M. (2016). *Pay It Forward: Investigating a Sustainable Model of Open Access Article Processing Charges for Large North American Research Institutions* (Final Report). University of California Libraries. https://escholarship.org/uc/item/8326n305

Springer Nature. (2025). *Spotlight on ten years of Transformative Agreements*. https://stories.springernature.com/oa-report-2024/spotlight-on-tas/

Steinberg, R. (2025). The business of transformative agreements. *The Journal of Academic Librarianship*, *51*(2), 103020. https://doi.org/10.1016/j.acalib.2025.103020

Sterman, L., McKelvey, H., & McLain, R. (2025). The Impact of Transformative Agreements on Reading and Publishing Behavior. *Journal of Open Initiatives in Academic Libraries*, *1*(1), 114–137. https://doi.org/10.58997/fx80dy21

Subbaraman, N. (2024, May 14). Flood of Fake Science Forces Multiple Journal Closures. *Wall Street Journal*. https://www.wsj.com/science/academic-studies-research-paper-mills-journals-publishing-f5a3d4bc

Swan, A., & Houghton, J. (2012). *Going for Gold? The costs and benefits of Gold Open Access for UK research institutions: further economic modelling* (Report to the UK Open Access Implementation Group). http://repository.jisc.ac.uk/610/2/Modelling_Gold_Open_Access_for_institutions_-_final_draft3.pdf

van Bellen, S., Alperin, J. P., & Larivière, V. (2025). Scholarly publishing's hidden diversity: How exclusive databases sustain the oligopoly of academic publishers. *PLOS ONE*, *20*(6), e0327015. https://doi.org/10.1371/journal.pone.0327015

Yahoo Finance. (2026). *Historical foreign exchange data*. https://finance.yahoo.com/

Zhang, L., Cao, Z., Shang, Y., Sivertsen, G., & Huang, Y. (2024). Missing institutions in OpenAlex: Possible reasons, implications, and solutions. *Scientometrics*, *129*(10), 5869–5891. https://doi.org/10.1007/s11192-023-04923-y

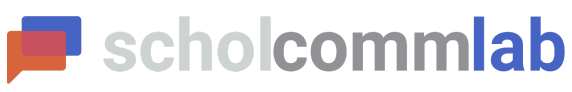

Zhang, L., Wei, Y., Huang, Y., & Sivertsen, G. (2022). Should open access lead to closed research? The trends towards paying to perform research. *Scientometrics*, *127*(12), 7653–7679. https://doi.org/10.1007/s11192-022-04407-5

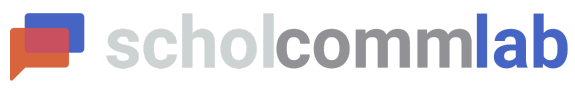

# Supplement 1: SQL query used to collect article counts from the OpenAlex snapshot

The SQL code used to query the OpenAlex snapshot and collect article counts per journal-year by OA type is:

```
SELECT
  APC_list.unique_id_version_2_1 AS unique_ID,
  APC_list.issn_l AS ISSN_L,
  APC_list.journal_normalized AS journal,
  publication_year AS year,
  works.primary_location.source.type,
  "ALL" as oa_type,
  COUNT(DISTINCT works.doi) AS DOIs,

FROM `subugoe-collaborative.openalex_walden.works` AS works

LEFT JOIN UNNEST(works.primary_location.source.issn) AS p_issn

INNER JOIN
  `subugoe-collaborative.resources.APC_list_v5` as APC_list
  ON p_issn = APC_list.issn_l

WHERE
  works.publication_year > 2018 AND works.publication_year < 2026 AND
  works.type = 'article'
  AND ( NOT REGEXP_CONTAINS(works.biblio.issue, '^[a-zA-Z]') OR works.biblio.issue IS
NULL ) -- remove issue starting with 'S', REGEXP_CONTAINS removes NULL automatically
so keep NULL issues
  AND ( NOT REGEXP_CONTAINS(works.biblio.issue, '^_') OR works.biblio.issue IS NULL )
-- remove issue starting with '_', keep NULL issues
  AND ( NOT REGEXP_CONTAINS(works.biblio.first_page, '^[sS]') OR
works.biblio.first_page IS NULL) -- remove first_page starting with 'S'
```

```
  AND ( NOT REGEXP_CONTAINS(works.biblio.first_page, '^_') OR works.biblio.first_page
IS NULL) -- or starting with underscore

  AND ( NOT (REGEXP_CONTAINS(works.title, '[0-9]{3} pp.')) )  -- remove book reviews
that have pattern in the title like '123 pp.'


GROUP BY unique_ID, ISSN_L, journal, works.primary_location.source.type, year



UNION ALL    -- keeps any duplicates



SELECT

  APC_list.unique_id_version_2_1 AS unique_ID,

  APC_list.issn_l AS ISSN_L,

  APC_list.journal_normalized AS journal,

  publication_year AS year,

  works.primary_location.source.type,

  works.open_access.oa_status AS oa_type,

  COUNT(DISTINCT works.doi) AS DOIs,


FROM `subugoe-collaborative.openalex_walden.works` AS works


LEFT JOIN UNNEST(works.primary_location.source.issn) AS p_issn


INNER JOIN

  `subugoe-collaborative.resources.APC_list_v5` as APC_list

  ON p_issn = APC_list.issn_l


WHERE

  works.publication_year > 2018 AND works.publication_year < 2026 AND

  works.type = 'article'

  AND ( NOT REGEXP_CONTAINS(works.biblio.issue, '^[a-zA-Z]') OR works.biblio.issue IS
NULL ) -- remove issue starting with 'S', REGEXP_CONTAINS removes NULL automatically
so keep NULL issues
```

```
  AND ( NOT REGEXP_CONTAINS(works.biblio.issue, '^_') OR works.biblio.issue IS NULL )
-- remove issue starting with '_', keep NULL issues

  AND ( NOT REGEXP_CONTAINS(works.biblio.first_page, '^[sS]') OR
works.biblio.first_page IS NULL) -- remove first_page starting with 'S'

  AND ( NOT REGEXP_CONTAINS(works.biblio.first_page, '^_') OR works.biblio.first_page
IS NULL) -- or starting with underscore

  AND ( NOT (REGEXP_CONTAINS(works.title, '[0-9]{3} pp.')) )  -- remove book reviews
that have pattern in the title like '123 pp.'


GROUP BY unique_ID, ISSN_L, journal, works.primary_location.source.type, year, oa_type


ORDER BY unique_ID, ISSN_L, year DESC, DOIs DESC
```

# Supplement 2: Supplementary tables and figures—unadjusted (nominal) estimates

The main text reports APC spend and revenue figures adjusted for inflation to 2025 USD using CPI Advanced Economies (see Methods). This supplement provides the corresponding unadjusted, nominal figures—the actual list-price values for each year, without adjustment. Table and figure numbers below correspond to their inflation-adjusted counterparts in the main text.

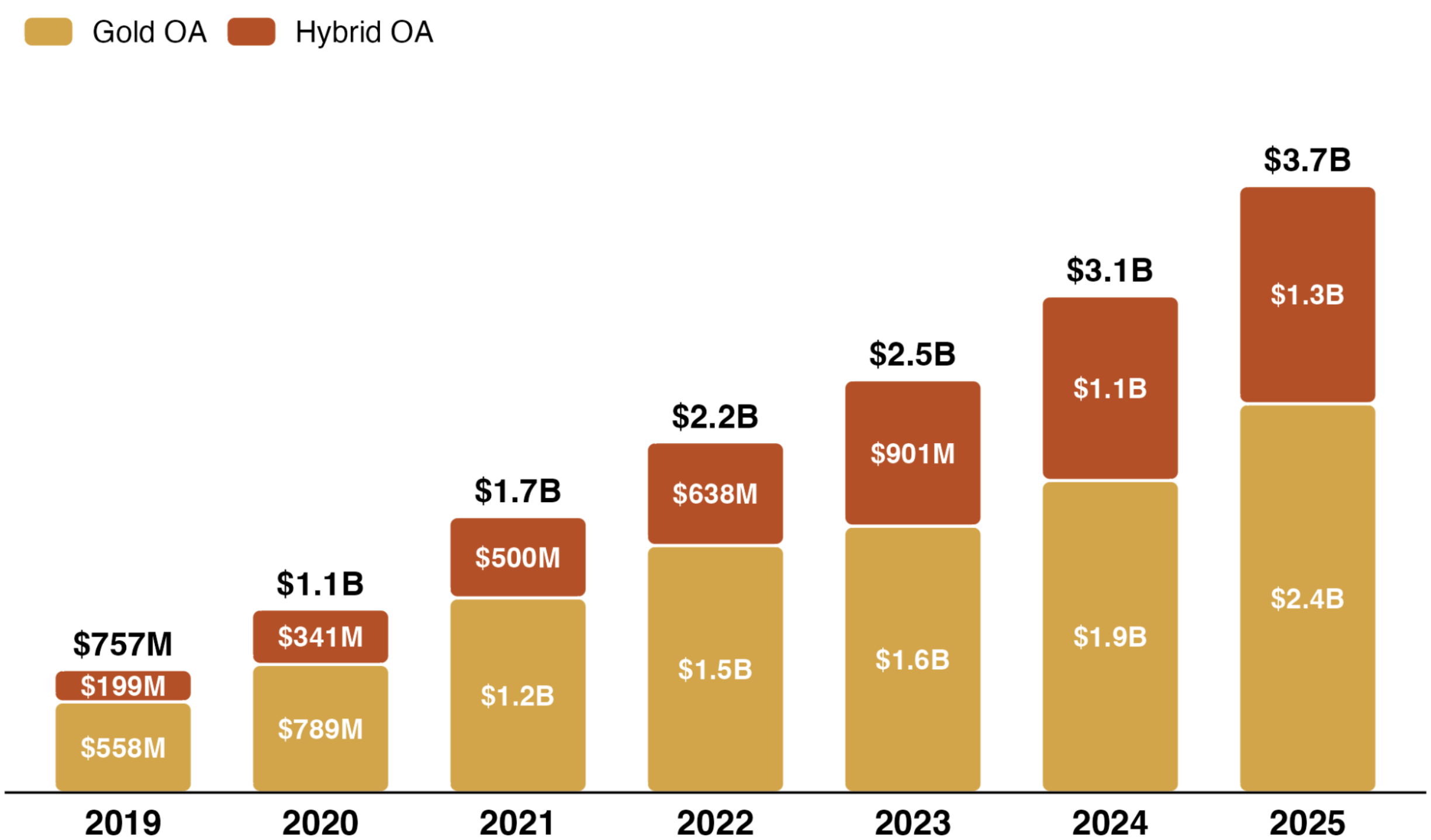

**Figure S2.2** Estimate of annual APC spend (in USD) per OA type.

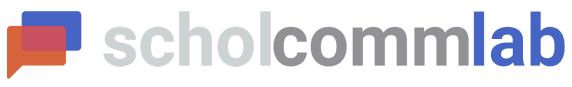

## Estimate of annual APC spend (in USD) by publisher

### Top 5 publishers

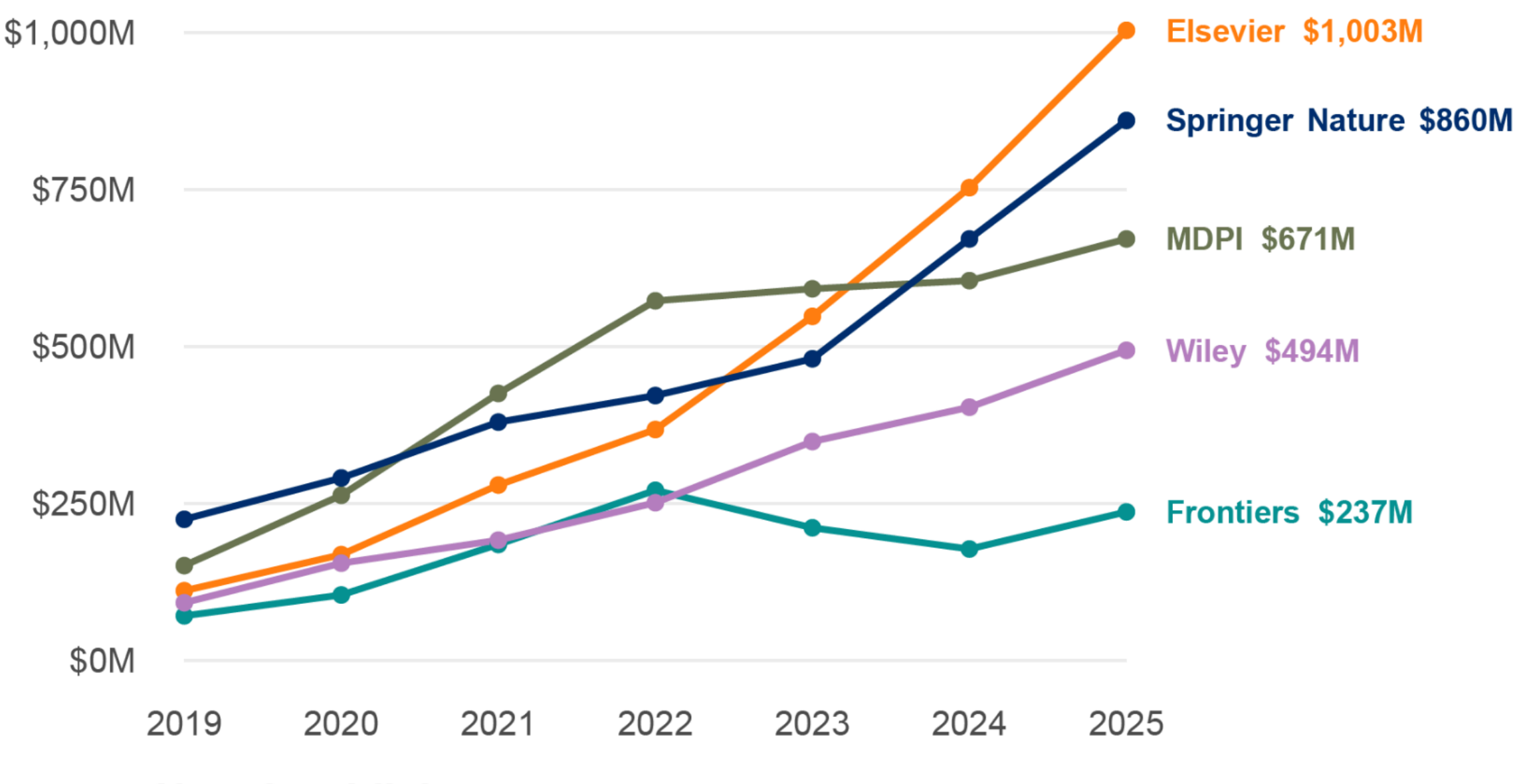

### Next 9 publishers

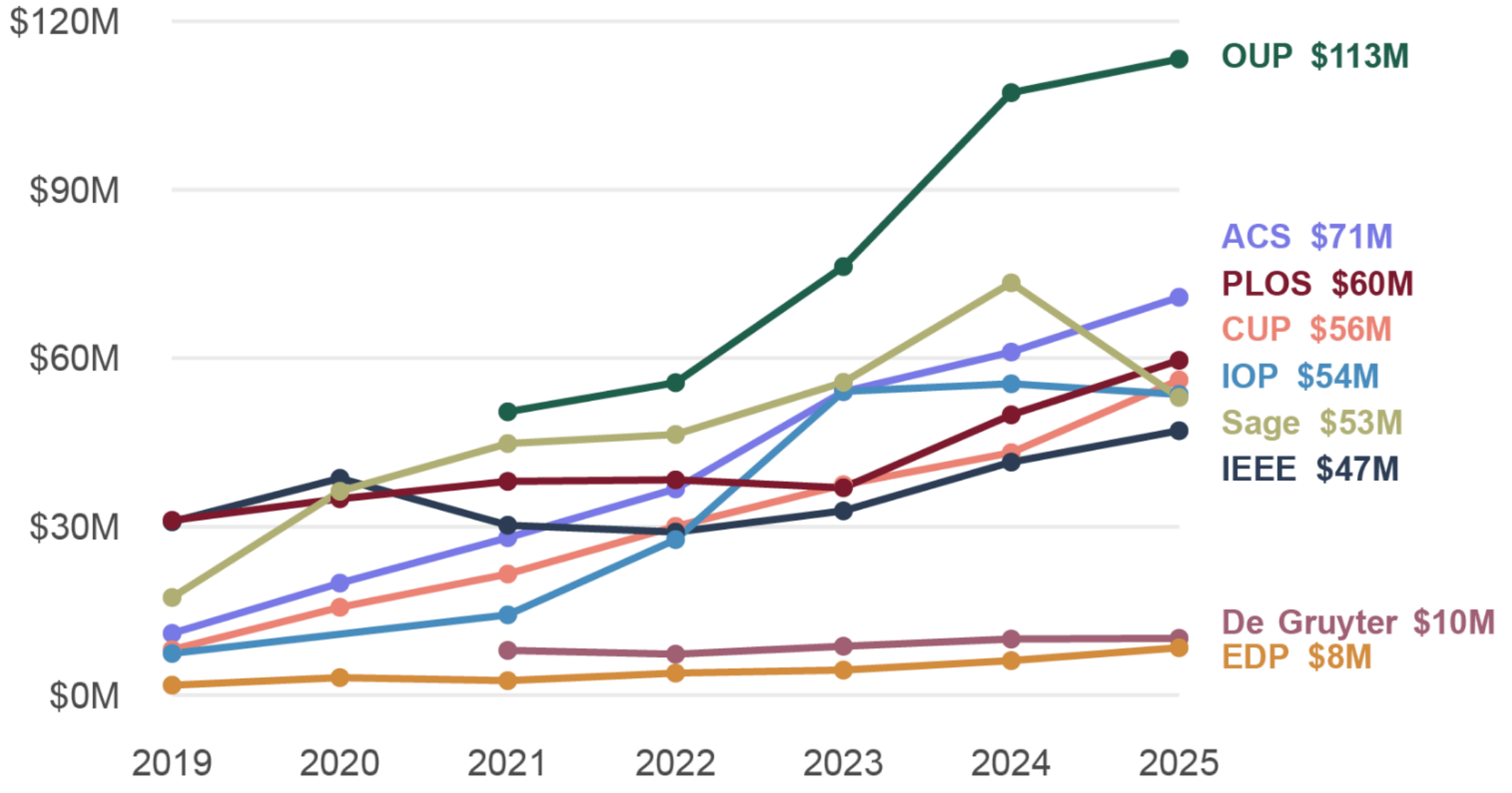

Sage's 2019 and 2025 figures reflect gold-only and hybrid-only data respectively.

**Figure S2.3** Estimate of annual APC spend (in USD) by publisher

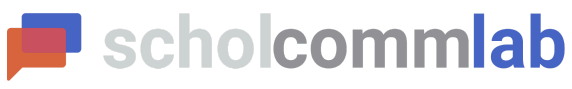

**Estimate of annual APC spend (in USD) by publisher and OA type**

Gold OA — Hybrid OA

Elsevier: $553M, $450M
Springer Nature: $575M, $285M
MDPI: $671M
Wiley: $281M, $213M
Frontiers: $237M
OUP: $60M, $54M
ACS: $58M, $13M
PLOS: $60M
CUP: $33M, $23M
IOP: $32M, $22M
Sage: $53M, $29M
IEEE: $38M, $9M
De Gruyter: $6M, $4M
EDP: $8M, $420K

Panels are ordered by total 2025 APC spend Note that the y-axis scale differs between panels.

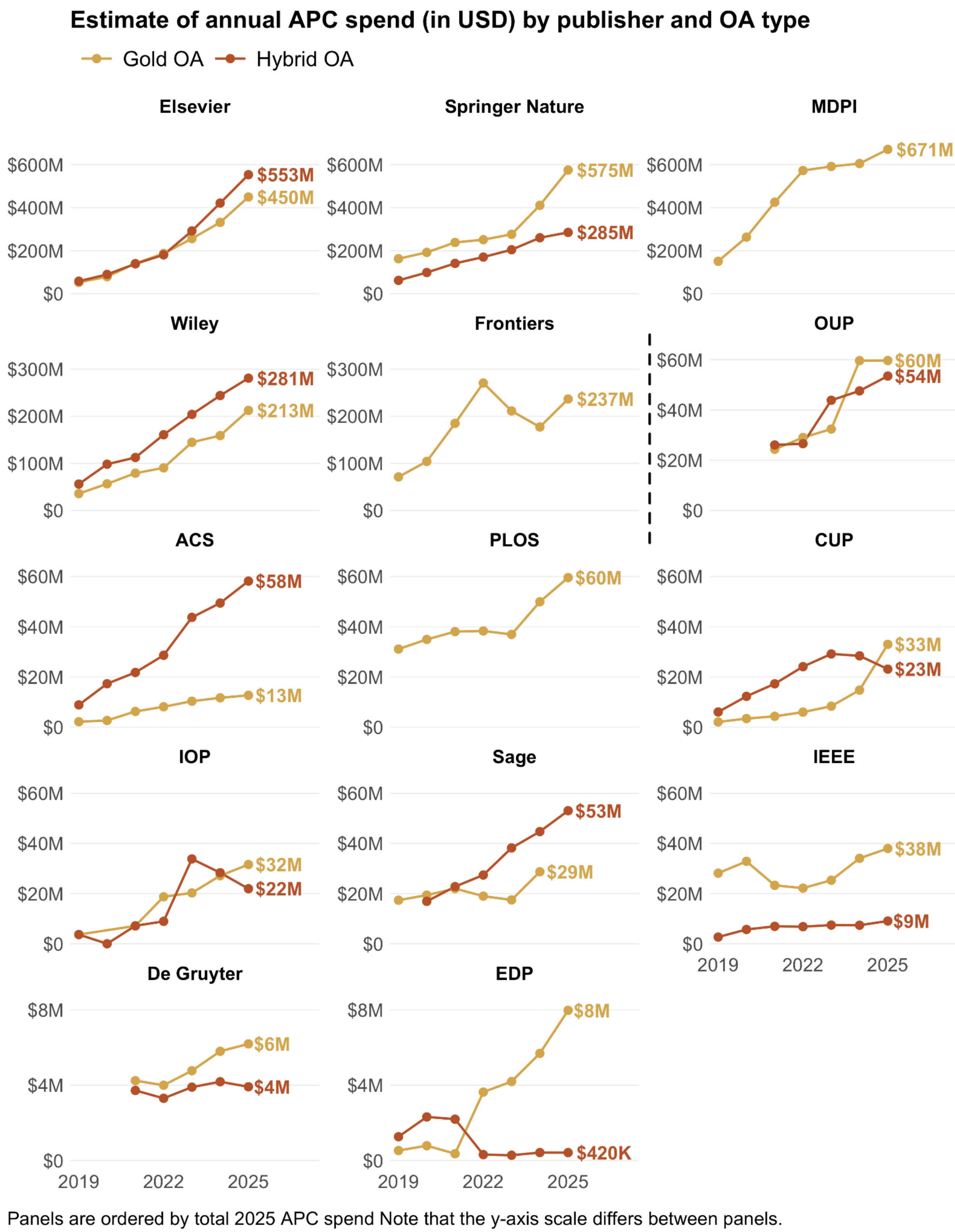

**Figure S2.4** Estimate of annual APC revenue (in USD) by publisher and OA type.

**Table S2.2** Growth of article output and APC spend (in USD) from 2019 to 2025 per publisher and OA type.

| 2019 to 2025 growth rate | Number of APC-able publications | | | Spend estimate based on APCs in USD | | |
|---|---|---|---|---|---|---|
| | Gold+Hybrid | Gold | Hybrid | Gold+Hybrid | Gold | Hybrid |
| **all publishers** | **+237.6%** | **+196.5%** | **+427.4%** | **+393.6%** | **+328.8%** | **+575.6%** |
| ACS | +326.9% | +133.9% | +629.6% | +544.6% | +491.9% | +557.4% |
| CUP | +372.4% | +648.3% | +199.1% | +592.6% | +1,499.4% | +282.9% |
| De Gruyter | n/a | n/a | n/a | n/a | n/a | n/a |
| EDP | +284.8% | +701.9% | -73.1% | +368.2% | +1,407.7% | -66.8% |
| Elsevier | +627.7% | +651.2% | +596.0% | +804.5% | +753.0% | +851.2% |
| Frontiers | +161.5% | +161.5% | n/a | +232.4% | +232.4% | n/a |
| IEEE | +27.2% | +15.8% | +176.9% | +52.6% | +35.0% | +237.5% |
| IOP | +422.9% | +475.3% | +357.2% | +623.8% | +752.5% | +494.7% |
| MDPI | +145.2% | +145.2% | n/a | +345.0% | +345.0% | n/a |
| OUP | n/a | n/a | n/a | n/a | n/a | n/a |
| PLOS | +30.6% | +30.6% | n/a | +91.6% | +91.6% | n/a |
| SAGE | n/a | n/a | n/a | n/a | n/a | n/a |
| Springer Nature | +172.3% | +150.6% | +255.7% | +282.5% | +253.2% | +359.5% |
| Wiley | +324.4% | +338.1% | +310.9% | +438.4% | +495.4% | +402.0% |

Note: n/a indicates that a growth rate could not be calculated. For De Gruyter, OUP, and Sage this reflects incomplete coverage—De Gruyter and OUP first appear in the dataset in 2021, and Sage's 2019 and 2025 records are gold-only and hybrid-only respectively. For Frontiers, MDPI, and PLOS, the absence of hybrid figures reflects their gold-only publishing model rather than a gap in the data.

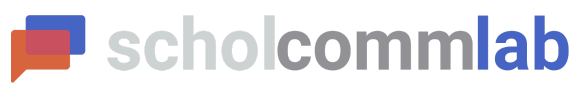

# Supplement 3: Publication counts

This supplement provides the number of APC-able publications underlying the spend estimates reported in the main text.

Table S3.1 Number of APC-able publications per OA type, 2019–2025.

| Publication year | Gold | Hybrid |
|---|---:|---:|
| 2019 | 293,735 | 63,506 |
| 2020 | 389,165 | 107,320 |
| 2021 | 562,126 | 150,099 |
| 2022 | 674,564 | 179,623 |
| 2023 | 685,441 | 245,595 |
| 2024 | 738,105 | 297,365 |
| 2025 | 871,024 | 334,954 |
| 2019-2025 | 4,214,160 | 1,378,462 |

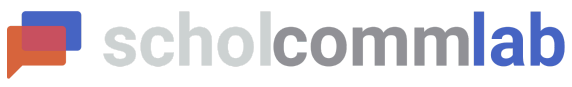

**Number of APC-able publications by OA type**

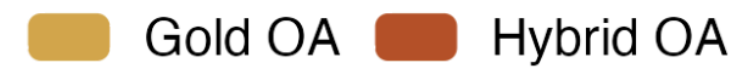

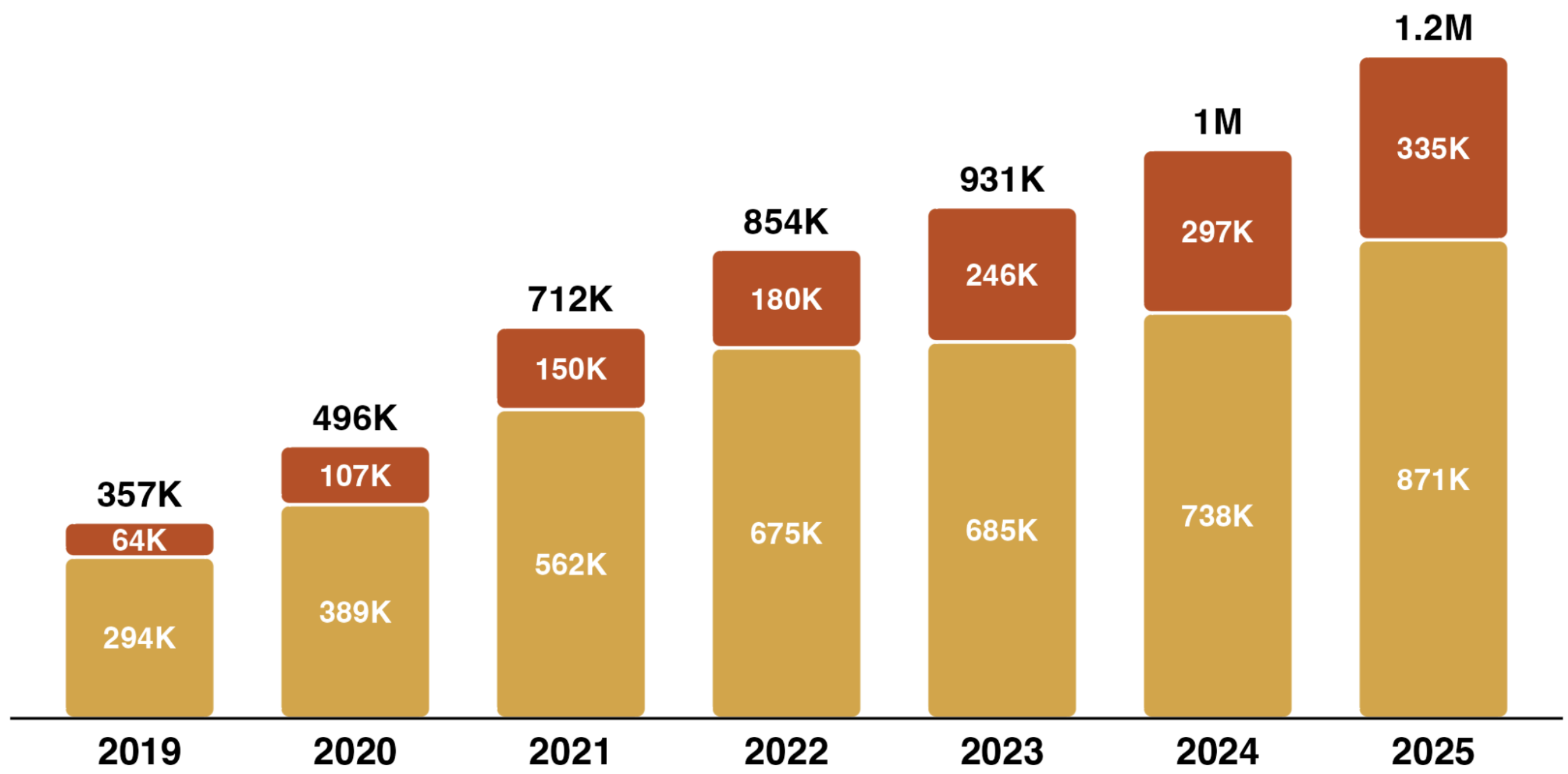

Some publishers are not covered in every year (De Gruyter and OUP begin in 2021, IOP is missing 2020, and Sage's 2019 and 2025 data are gold-only and hybrid-only respectively).

Figure S3.2 Number of annual APC-able publications by OA type, 2019–2025.

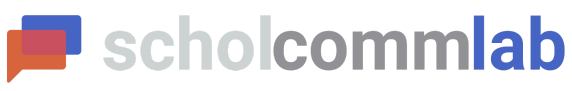

## Number of APC-able publications by publisher

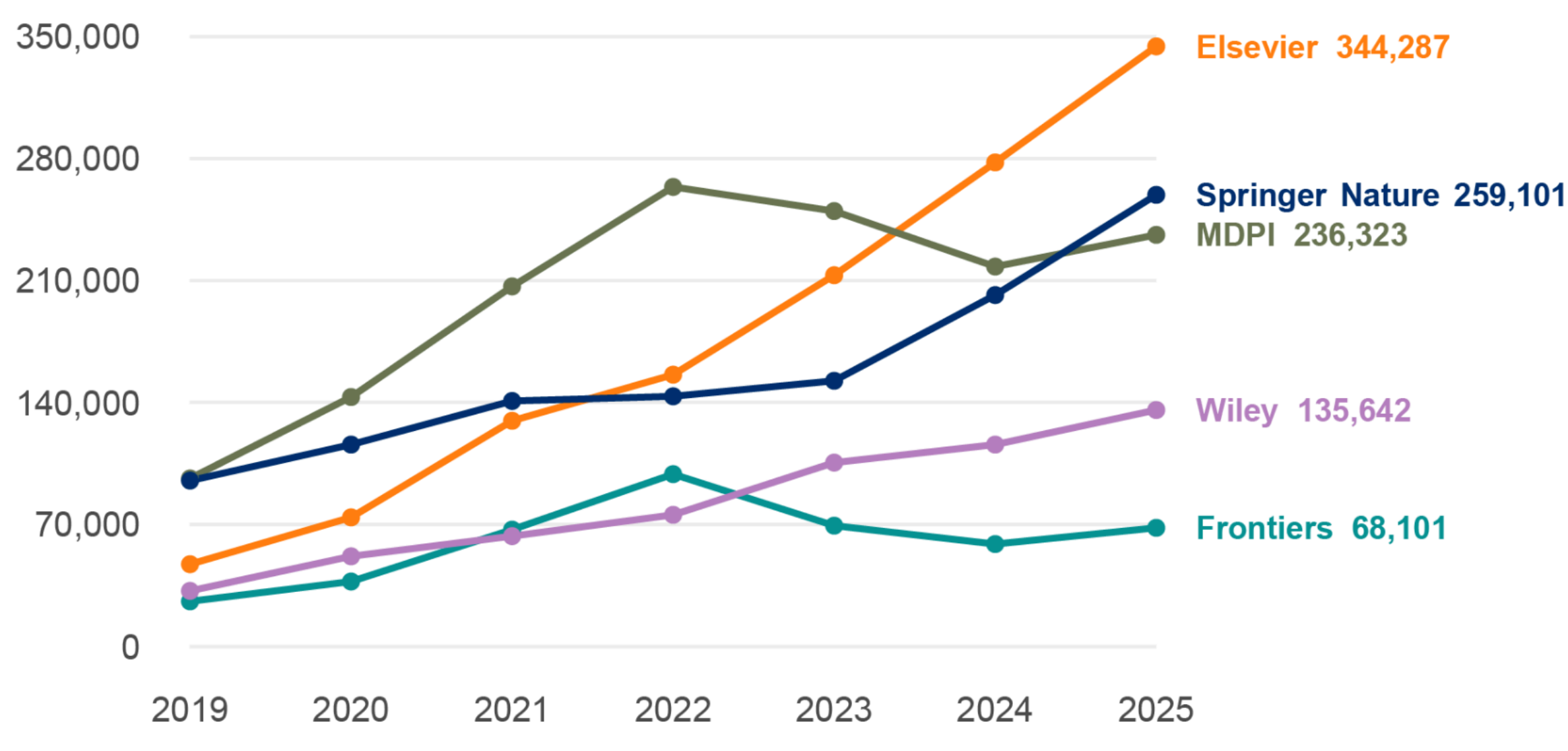

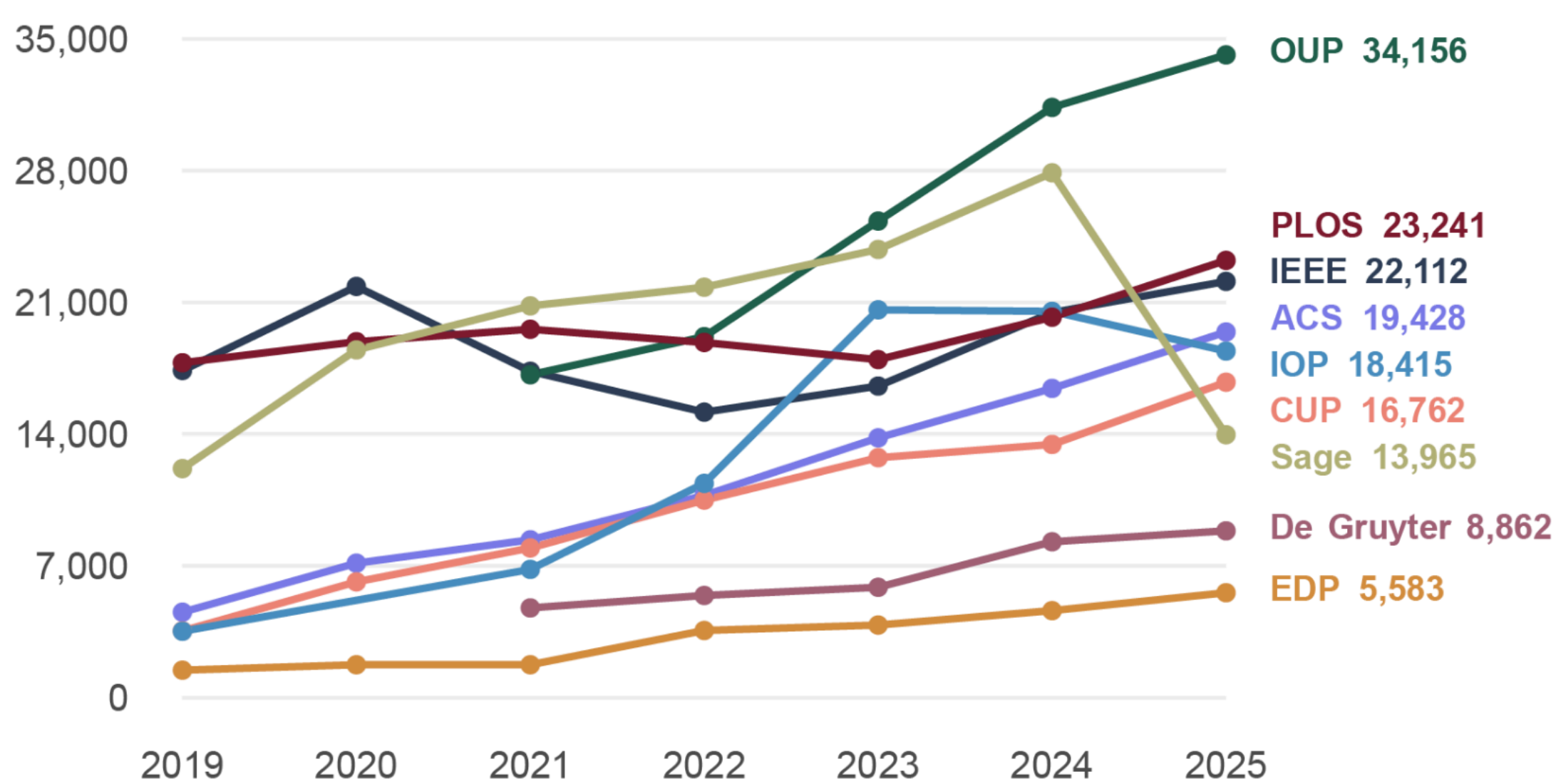

Sage's 2019 and 2025 figures reflect gold-only and hybrid-only data respectively.

Figure S3.3 Number of APC-able publications by publisher, 2019–2025.

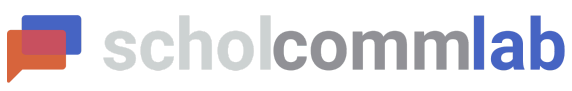

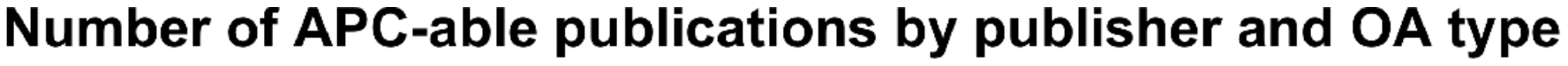

Gold OA  Hybrid OA

Elsevier: 204K, 140K
Springer Nature: 189K, 70K
MDPI: 236K
Wiley: 69K, 66K
Frontiers: 68K
OUP: 21K, 13K
PLOS: 23K
IEEE: 19K, 3K
ACS: 13K, 6K
IOP: 11K, 7K
CUP: 10K, 7K
Sage: 16K, 14K
De Gruyter: 7K, 2K
EDP: 5K, 210

2019 2022 2025

Panels are ordered by total number of APC-able publications. Note that the y-axis scale differs between panels.

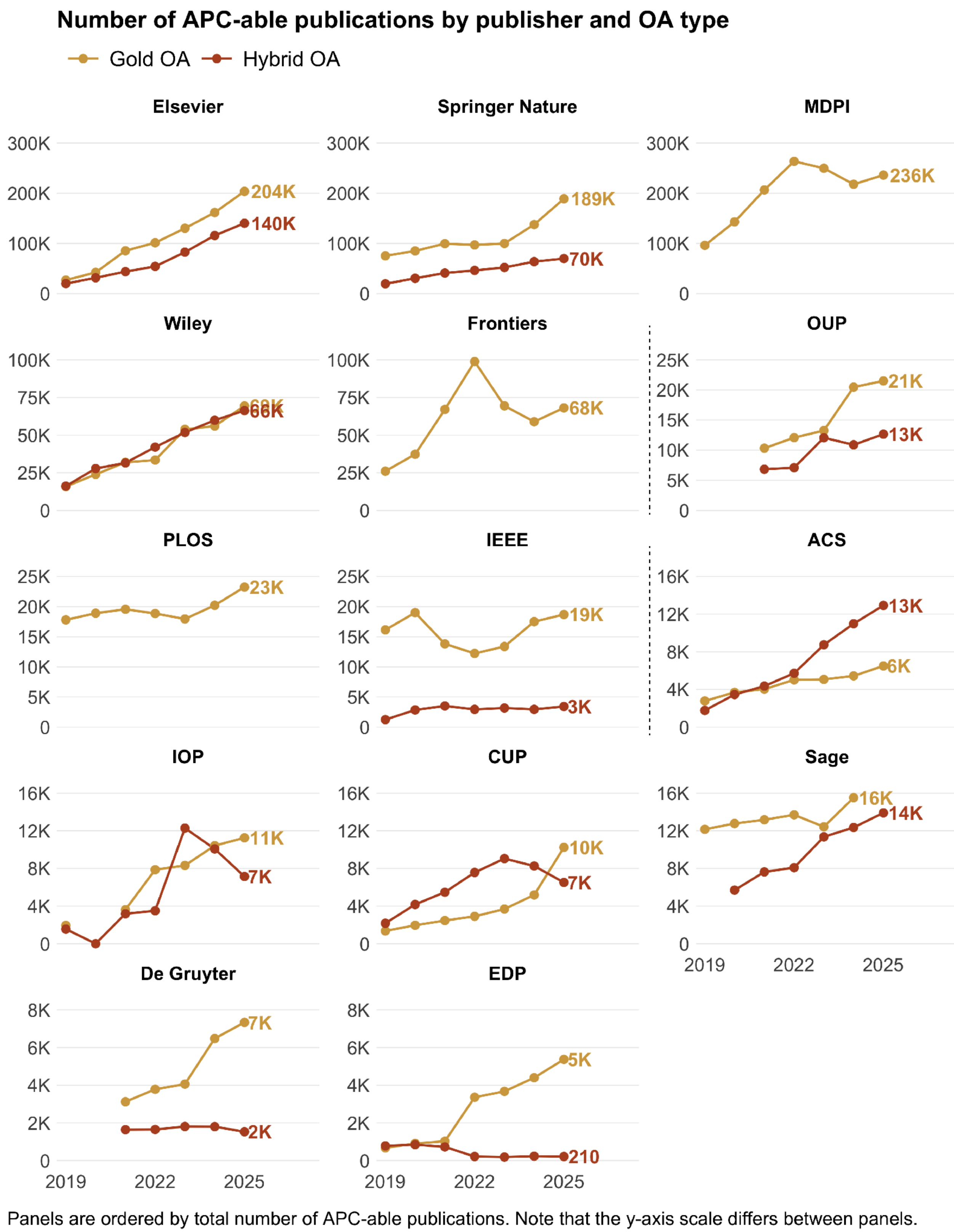

Figure S3.4 Number of APC-able publications by publisher and OA type, 2019–2025.